\documentclass[%
 reprint,
superscriptaddress,
 amsmath,amssymb,amsfonts,
 aps,
 prl,
 floatfix,
 longbibliography,nobibnotes
]{revtex4-2}

\usepackage[utf8]{inputenc}

\usepackage[pdftex]{graphicx} \graphicspath{{}}
\usepackage{float} 
\usepackage{xcolor}
\usepackage{verbatim}
\usepackage{soul}
\usepackage[pdftex,colorlinks=true]{hyperref}
\hypersetup{
  colorlinks=true,
  linkcolor=blue,
  urlcolor=cyan,
}

\usepackage{bm}

\usepackage{mathtools}

\DeclarePairedDelimiterX\braket[2]{\langle}{\rangle}{#1 \delimsize\vert #2}
\DeclarePairedDelimiterX\expval[3]{\langle}{\rangle}{#1 \delimsize\vert #2  \delimsize\vert #3}
\DeclarePairedDelimiterX\proj[2]{\delimsize\vert#1\rangle}{\langle#2\delimsize\vert}{ }

\usepackage{microtype}

\begin{document}

\title{Nonequilibrium Critical Scaling of a Squeezing Phase Transition}

\author{Arman Duha}
\affiliation{Department of Physics, Oklahoma State University, Stillwater, Oklahoma 74078, USA}

\author{Samuel E. Begg}
\affiliation{Department of Physics, Oklahoma State University, Stillwater, Oklahoma 74078, USA}

\author{Thomas Bilitewski}
\email{thomas.bilitewski@okstate.edu}
\affiliation{Department of Physics, Oklahoma State University, Stillwater, Oklahoma 74078, USA}

\date{\today}

\begin{abstract}
We investigate phase transitions in the nonequilibrium dynamics of power-law interacting spin-1/2 bilayer XXZ models, which have recently been shown to allow generation of entanglement in the form of two-mode squeezing. We find a transition between a collective phase characterized by Heisenberg limited squeezing and a partially collective phase with scalable squeezing. We identify universal scaling of the squeezing dynamics in terms of system parameters and a divergent time-scale, establishing these as distinct dynamical phases within the framework of non-equilibrium critical phenomena. Our work demonstrates a novel dynamical phase transition with potential applications in quantum sensing and quantum simulation in cold-atomic, molecular or Rydberg platforms.
\end{abstract}
\maketitle
%
{\it Introduction---}%
Matter in equilibrium can be classified into phases characterized by universal behavior irrespective of microscopic details \cite{Zinn-Justin2002,Goldenfeld2019,RevModPhys.49.435,Tauber_2014}, which facilitates understanding of large classes of systems. 
Similarly, the dynamics of systems out-of-equilibrium can present universal properties \cite{Tauber_2014,LUBECK_2004,Berges2015} allowing a classification of non-equilibrium phases of matter \cite{Hinrichsen2000,Odor2004}. This is well understood in the classical case, with paradigmatic examples in flocking  \cite{Vicsek1995,Toner1995,Benoit2019}, Kardar-Parisi-Zhang physics \cite{Kardar1986}, aging \cite{Calabrese_2005}, and directed percolation \cite{Hinrichsen2000}. Nonequilibrium universality has also been established in quantum systems, e.g. in the form of dynamical quantum phase transitions \cite{PhysRevLett.97.200601,Heyl2013,Heyl2017,PhysRevLett.120.130601}, measurement-induced phase transitions \cite{Skinner2019,Fisher2019}, and in pre-thermal and driven phases of matter \cite{Else_2017,Yao_2020,sieberer2023universalitydrivenopenquantum}. %

An important nonequilibrium process \cite{RevModPhys.83.863} is the generation of entangled states for quantum metrology \cite{RevModPhys.89.035002,RevMod_Metrology_2018,montenegro2024reviewquantummetrologysensing} via time evolution from simple, easy to prepare, initial states. This includes spin-squeezed states \cite{Wineland1992,wineland1994,Kitagawa1993} with spin projection noise below the standard quantum limit enabling quantum enhanced sensing. Spin squeezing has been predicted to occur in power law interacting spin systems \cite{fossfeig2016entanglementspinsqueezinginfiniterangeinteractions,PhysRevLett.125.223401,PhysRevA.109.L061304,Block_2024,Roscilde2022,Roscilde2023,Roscilde2024,koyluoglu2025squeezingheisenberglimitlocally} and recently demonstrated in a number of experiments \cite{Eckner2023,Franke2023a,Bornet2023a,Hines2023}. The wide range of experimental platforms realizing long-range interactions \cite{RevModPhys.95.035002}, from Rydberg \cite{Browaeys2020,Saffman2010,Morgado2021} and magnetic atoms \cite{Chomaz2023},  over polar molecules \cite{Baranov2012,Bohn2017,Moses2017} and trapped ions \cite{Blatt2012, Monroe2021}, to cavity systems \cite{Norcia2018,Davis2019,Mivehvar2021}, makes them particularly promising candidates  to controllably investigate many-body nonequilibrium dynamics and entanglement generation. %

A natural question is whether the dynamical preparation of  highly entangled squeezed states can be understood through the lens of phases and universality. 
Recent works have demonstrated that spin squeezing may be related to equilibrium phases of matter \cite{Block_2024,Roscilde2024}. %
Specifically, Block et al. \cite{Block_2024} showed that quenches within the easy plane ferromagnetic phase of quantum magnets can exhibit scalable squeezing in their time dynamics. %
It was then shown that a similar picture emerges even for short-range interacting systems with only quasi-long-range order \cite{Roscilde2024}, with critical slowing down protecting scalable squeezing. 

\begin{figure}[t]
\includegraphics[width=\columnwidth]{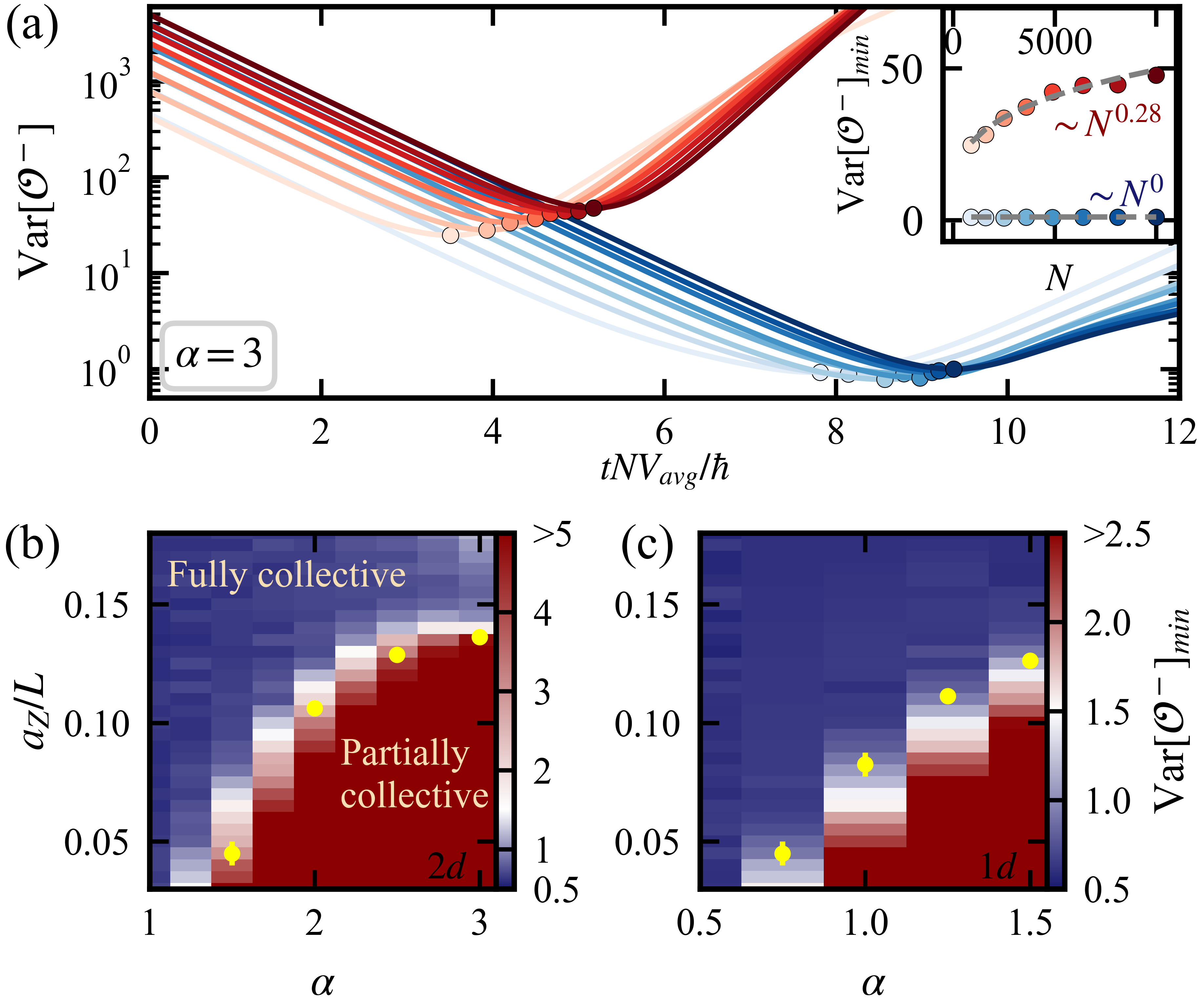}
\caption{Dynamical phases of squeezing. (a) Time evolution of the variance of the squeezed quadrature ${\rm Var}[\hat{\mathcal{O}}^-] $ in the fully collective regime (blue, $a_Z/L=0.17$) and partially collective regime (red, $a_Z/L=0.04$) for $\alpha = 3$ in $2d$. The opacity increases with system size $N \in \{9\times 10^2,10^4\}$. %
Inset: System size scaling of the minimum variance ${\rm Var}[\hat{\mathcal{O}}^-]_{\rm min}$ defining the fully, ${\rm Var}[\hat{\mathcal{O}}^-]_{\rm min}\sim N^0$, and partially collective regime $\sim N^{0.28}$. 
(b, c) Dynamical phase diagram as a function of power law exponent $\alpha$ and aspect ratio $a_Z/L$ in (b) $2d$ and (c) $1d$. The fully collective regime spans all $a_Z/L$ for $\alpha \leq d/2$, whereas a transition occurs at a critical $a_Z/L$ value (yellow marker) when $\alpha>d/2$.  
\label{fig:fig1}}
\end{figure} 

Here, we demonstrate non-equilibrium critical scaling in the full time dynamics of a spin system that exhibits two mode squeezing. We first establish distinct dynamical regimes in terms of the scaling of squeezing with system size. %
We explain the existence of the fully collective phase and the transition based on the Bogoliubov excitation spectrum. %
We then demonstrate the phenomenology of critical phenomena by resolving critical exponents governing the scaling with system-size and divergent time-scales as a function of control parameters. The exponents are consistent (within error) for different lattices; this independence from microscopics is highly suggestive of an underlying nonequilibrium universality. We further directly connect the observed system size scaling of squeezing to the universal scaling behavior.

{\it Model---}%
We consider power law interacting spin 1/2 systems in ladders with dimension $d=1$ or square bi-layers with dimension $d=2$ described by
\begin{equation}
\hat{H} = 1/2 \sum_{\eta} \sum_{\boldsymbol{i},\boldsymbol{j} \in \eta} V_{\boldsymbol{i}\boldsymbol{j}} \, \vec{s}_{\boldsymbol{i}} \cdot \vec{s}_{\boldsymbol{j}} +  \sum_{\mathclap{\boldsymbol{i} \in A, \boldsymbol{j} \in B}} \, V_{\boldsymbol{i}\boldsymbol{j}} \, (\hat{s}_{\boldsymbol{i}}^x \hat{s}_{\boldsymbol{j}}^x +\hat{s}_{\boldsymbol{i}}^y \hat{s}_{\boldsymbol{j}}^y ) \label{eq:model}
\end{equation}
where the spin operators $\hat{s}_{\boldsymbol{i}}^{\mu}=\hat{\sigma}_{\boldsymbol{i}}^{\mu}/2$, with the  Pauli matrices $\hat{\sigma}^{\mu}$, act on the spin at site $\boldsymbol{i}$, specifying the position in layers (denoted by $\eta = A, B$) of a one-dimensional spin-ladder or a two-dimensional bi-layer. 
 
The chains or layers are separated by a distance $a_Z$, which we measure in terms of the in-layer spacing $a_{\mathrm{lat}}$, set to 1 in the following. %
We consider power-law interactions with exponent $\alpha$ of the form $V_{\boldsymbol{i}\boldsymbol{j}} = |\boldsymbol{r_{i}} - \boldsymbol{r_{j}}|^{-\alpha}$, modeling a range of experimental platforms \cite{RevModPhys.95.035002} from Rydberg atoms ($\alpha=3,6$) \cite{Browaeys2020,Saffman2010,Morgado2021}, polar molecules ($\alpha=3$)\cite{Baranov2012,Bohn2017,Moses2017}, magnetic atoms ($\alpha=3$)\cite{Chomaz2023}, trapped ions ($0<\alpha<3$) \cite{Blatt2012, Monroe2021}, or cavity systems ($\alpha=0$) \cite{Norcia2018,Davis2019,Mivehvar2021}. %
The spin structure of the interactions in Eq. \ref{eq:model} can be engineered from Ising interactions \cite{PhysRevA.109.L061304} using Floquet techniques \cite{Lukin_2020_Robust}, which has been demonstrated across experimental platforms \cite{Lukin_2020_Robust,Geier2021,Scholl2022,Christakis2023,Miller2024,PRXQuantum.4.010334}.

We initialize non-equilibrium dynamics from an initial state of oppositely polarized layers, $\langle\hat{S}_A \rangle= - \langle \hat{S}_B\rangle = N/2 \, \hat{z}$ \footnote{Noting that this initial state is an infinite temperature state of the inter-layer interactions it would not be expected to show equilibrium order}. The interlayer spin-exchange interactions, second term of Eq.~\ref{eq:model}, will dynamically create entangled pairs of excitations and two-mode squeezing \cite{PhysRevA.109.L061304,Bilitewski2023a,Bilitewski2023}. Performing a Holstein-Primakoff transformation (HPT) \cite{Holstein1940} of the total layer spins $\hat{S}^{\mu}_{A/B} = \sum_{\bm{i} \in A/B} \ \hat{s}_{\boldsymbol{i}}^{\mu}$ we obtain $ H \approx N V_{\mathrm{avg}}  (\hat{a}^{\dagger} \hat{b}^{\dagger} + \hat{a} \hat{b})/2$, the two-mode squeezing Hamiltonian \cite{Agarwal2013,Schumaker1985,Caves1985}, where $V_{\mathrm{avg}}$ is the average interlayer interaction and $\hat{a}$ ($\hat{b}$) describes excitations in layer A (B). This predicts exponential growth/reduction of variances $\mathrm{Var}[\mathcal{O}^{\pm}] =N/2 \, e^{\pm N V_{\mathrm{avg}}t/\hbar}$, i.e. squeezing of $\mathcal{O}^{-}=S^x_A + S^y_B$ or $S^y_A-S^x_B$, and anti-squeezing of $\mathcal{O}^{+}=S^x_A - S^y_B$ or $S^y_A + S^x_B$ \cite{supplemental}. We emphasize the mechanism for generating spin-squeezing extends beyond the applicability of the HPT in the full spin model.

{\it Dynamical transitions in spin squeezing---}%
We now explore the dynamical regimes of the model Eq.~\ref{eq:model} as a function of the dimensionality of the system $d$, power-law exponents of the interactions $\alpha$, and aspect ratio $a_Z/L$ using the discrete truncated Wigner approximation (dTWA) \cite{Schachenmayer2015,Zhu_NewJournalofPhysics_21_2019,Muleady_PRL_2023}.
%
DTWA, while semi-classical, has been shown to yield near exact results for powerlaw spin systems \cite{Muleady_PRL_2023}, and compares well to exact dynamics for our model \cite{PhysRevA.109.L061304}, but may fail to capture long-time and thermalisation dynamics in particular for short-range models \cite{Roscilde2024}.

 Our main results are summarized in Fig.~\ref{fig:fig1}. Panel (a) shows the time evolution of the variance of the squeezed quadratures, $\mathrm{Var}[\mathcal{O^-}]$, for $\alpha=3$ with $a_Z/L=0.04$ (red) and $a_Z/L=0.17$ (blue). In both cases the spin dynamics follow the two-mode squeezing prediction of exponentially decreasing variance up to a saturation point of minimal variance. Results are shown for a range of system sizes (opacity), and indicate that for the larger (smaller) $a_Z/L$ value this minimum does not (does) depend on the system size as shown directly in the inset of Fig.~\ref{fig:fig1}(a). The results for smaller $a_Z/L$ approximately follow $\mathrm{Var}[\mathcal{O}^-]_{\rm min}\sim N^{p}$ with $p = 0.28$, whereas the larger $a_Z/L$ data exhibits $p = 0$. This suggests the presence of two dynamical regimes: a partially collective regime (red) with $p > 0$ and a fully collective regime (blue) with $p = 0$.
 
The reduced variance directly results in an improved sensitivity of measuring a phase of rotation $\phi$ around $\hat{S}^z_A-\hat{S}^z_B$  as $(\Delta \phi)^2 =\frac{(\Delta \mathcal{O})^2}{ (\langle \hat{S}_A^z - \hat{S}_B^z \rangle)^2}$\cite{PhysRevA.109.L061304,Sundar2023,supplemental}. Noting that the polarization in all cases considered here remains of order $N$ at the time of optimal squeezing (SM \cite{supplemental}) the sensitivity scales as $N^{p-1}$ achieving Heisenberg limited scaling for $p=0$, and enabling quantum enhanced sensing for $p<1$, at a time of order $\log(N^{p})/(N V_{\mathrm{avg}})$.
 
We map out the two dynamical regimes in Fig.~\ref{fig:fig1}(b) and (c) as a function of the aspect ratio $a_Z/L$ and power-law exponents $\alpha$ for $1d$ (b) and $2d$ (c). For $\alpha > d/2$,
a critical $a_Z/L$ exists (markers), above which the minimal variance is of order $1$ (blue), signifying the fully collective regime. In contrast, below the critical value the minimal variance is greater than $\mathcal{O}(1)$ and scales with system size, marking the departure from the fully collective regime into a partially collective regime. The dynamics remain fully collective if $\alpha \leq d/2$. The transition points $(a_Z/L)_*$ (yellow markers) in Fig.~\ref{fig:fig1} (b) and (c) are defined by the scaling exponent $p$ becoming non-zero, see Fig.~S11 \cite{supplemental}.

{\it Bogoliubov Analysis---}%
\begin{figure}[t]
\includegraphics[width=\columnwidth]{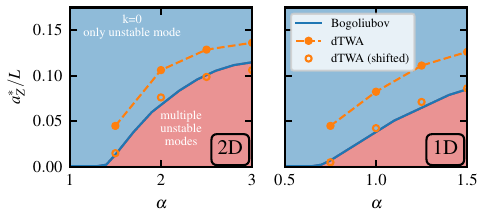}
\caption{Comparison of phase-boundary within Bogoliubov analysis and dTWA. Critical aspect ratio $(a_Z/L)_*$ (Bogoliubov, solid line) as a function of powerlaw exponent $\alpha$ separating the region, in which only the $k=0$ mode is unstable (shaded blue), and the region with multiple unstable modes. Compared with dTWA results (markers, dashed line) (Fig~1(b,c)).
\label{fig:fig_Bog_3}}
\end{figure} 
We now demonstrate how the existence of the fully collective phase and location of the transition can be understood from Bogoliubov theory.

Within quadratic order the Hamiltonian, Eq.~\ref{eq:model}, can be diagonalised in momentum space \cite{supplemental,Bilitewski2023,Dominguez2024}, resulting in quasi-energies of the form
$
\sqrt{\varepsilon_{\bm{k}}^2 -|\Omega_{\bm{k}}|^2}
$
with  $\varepsilon_{\bm{k}} = \epsilon_{\bm{k}} - \epsilon_{0}$, where
$
\epsilon_{\bm{k}} = \frac{1}{L^{d}} \sum_{\bm{j} } e^{-i \bm{j} \bm{k}}  V_{\bm{j}}
$
and
$
\Omega_{\bm{k}} = \frac{1}{L^{d}} \sum_{\bm{j}\in {A - B}  } e^{-i\bm{j}\bm{k}}  V_{\bm{j}}
$
are the Fourier-transforms of the intra- and inter-layer interactions respectively. If $|\Omega_{\mathbf{k}_{c}}|>|\varepsilon_{\mathbf{k}_{c}}|$ for a quasi-momentum $\mathbf{k}_{c}$, the eigenenergies are imaginary resulting in exponential growth of the corresponding momentum modes. Critically, this is always the case for $\textbf{k}=0$ as a consequence of the Heisenberg in-plane and XX inter-plane interactions, resulting in the two-mode squeezing dynamics observed in Fig.~\ref{fig:fig1}(a) \cite{supplemental}.

Generically, there will be a region in momentum space of unstable modes. We may therefore distinguish two parameter regimes of the model. %
%
In one, only the $k=0$ mode is unstable and is predicted to grow exponentially to generate entanglement in the form of two-mode squeezing, while all other momentum modes are stable and remain at low population. This corresponds to the fully collective squeezing phase observed in Fig.~\ref{fig:fig1}.
In contrast, in the regime with multiple unstable modes, the exponential growth of finite k-modes (if it were to continue unabated) would rapidly reduce the collective spin length and depolarise the system \cite{supplemental}. %
Thus, the quadratic Bogoliubov theory predicts the existence of a fully collective regime, with a transition into a non-squeezing phase. 

We compare in Fig.~\ref{fig:fig_Bog_3} the phase diagram obtained from Bogoliubov theory with the phase-boundary obtained from dTWA simulations. While the transition point is not quantitatively captured in the Bogoliubov results (solid line), we observe good qualitative agreement in the shape of the transition line between dTWA (filled markers) and shifted dTWA results (empty markers). %
This agreement indicates that indeed the absence of finite momentum unstable modes is related to the existence of the fully collective phase. %
Moreover, the partially collective phase with scalable squeezing exists in the region predicted to have multiple unstable modes, requiring interactions to suppress the generation of finite momentum excitations, making it a genuinely interacting phenomenon beyond basic quadratic Bogoliubov theory \cite{supplemental}.

{\it Scaling of variance minima---}%
\begin{figure}[t]
\includegraphics[width=\columnwidth]{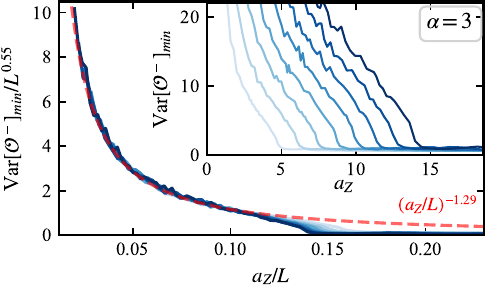}
\caption{Universal scaling collapse of minimal squeezing.
Inset: Dependence of minimal variance $\mathrm{Var}[\mathcal{O}^-]_{\rm min}$ on layer spacing $a_Z$, for different system sizes ($L = 30$ to 100) with $\alpha =3$ in $2d$. The opacity increases with system size. Main panel: Rescaled minimal variance $\mathrm{Var}[\mathcal{O}^-]_{\rm min}/L^{0.55}$ versus aspect ratio $a_Z/L$ for the same system sizes, demonstrating a universal collapse onto a single curve. Red dashed line power-law $(a_Z/L)^{-1.29}$ obtained from the scaling ansatz, Eq.~\ref{eq:minscale}. \label{fig:fig2}}
\end{figure}
We now characterize the respective phases in terms of a scaling collapse of the minimal variance. The inset of Fig.~\ref{fig:fig2} shows the minimal variance versus layer separation $a_Z$ for different system sizes, $\alpha = 3$ in $2d$, clearly showing that the critical $a_Z$ scales with $L$. %
In the main panel of Fig.~\ref{fig:fig2} we plot the data as a function of $a_Z/L$, which corresponds to a vertical cut of the phase diagram in Fig. \ref{fig:fig1}(b), and rescale the minimal variance by $L^{-0.55}$, equivalent to $N^{0.28}$ observed in the inset of Fig. \ref{fig:fig1}(a) since $N = L^2$ in $2d$. Under this rescaling the data collapses to a single curve, or scaling function, demonstrating that the aspect ratio $a_Z/L$ controls the transition, and that both regimes show scaling of the minimal variance characterized by a single distinct $p$ value. %
The dashed line $(a_Z/L)^{d_V}$, with $d_V = -1.29$, is an excellent fit to the scaling function which we will later demonstrate to be a critical exponent. Details of the calculation of scaling exponents and extended data can be found in the Supplemental Material (SM) \cite{supplemental}.

{\it Critical scaling in the partially collective phase---}%
\begin{figure}[t]
\includegraphics[width=\columnwidth]{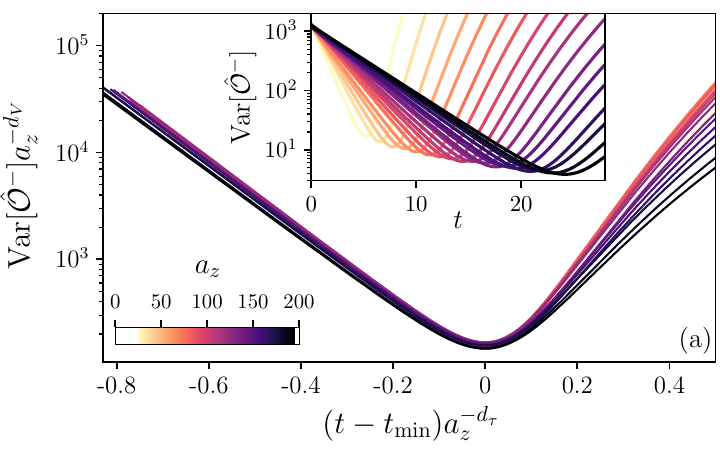}
\includegraphics[width=\columnwidth]{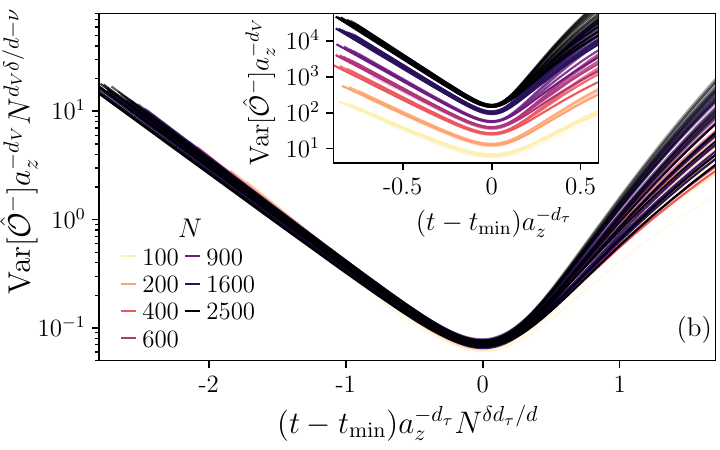}
\caption{Universality of the partially collective phase. (a) ${\rm Var}[\hat{\mathcal{O}}^{-}] a_Z^{-d_{V}}$ vs rescaled time $t a_Z^{-d_{\tau}}$ for different $a_Z$ values in the partially collective regime (colorbar). Inset: raw data ${\rm Var}[\hat{\mathcal{O}}^{-}]$ vs time $t$  for a 1d system with $\alpha = 1.5$.  (b) 
Same as in (a) but for a range of system sizes (legend) and with  additional system size rescaling, giving variance $ {\rm Var}[\hat{\mathcal{O}}^{-}]N^{-\nu}a_{Z,\delta}^{-d_{V}}$ vs rescaled time $(t-t_{\rm min}) a_{Z,\delta}^{-d_{\tau}}$. For each $N$, a range of $a_Z$ values are plotted with increased fading for smaller values. Inset: data without $N$ rescaling. The exponents are $d_{\tau} = 0.63$, $d_V = -0.69$, $\nu  = 0.81$ and $\delta=0.25.$}
\label{fig:timecollapse}
\end{figure} 
Having demonstrated a clear transition between two regimes, quantified by the exponent $p$ which plays the role of an order parameter, 
we now turn to demonstrating universal scaling in the vicinity of the critical point, just as occurs for equilibrium phases.

Since the variance minima occur at different times $t_{\rm min}$ that grow with increasing $a_Z$, the scaling collapse in Fig. \ref{fig:fig2} is suggestive that the partially collective phase may obey a universal scaling that depends on a diverging relaxation time-scale, in analogy to critical slowing down at equilibrium transitions. We now show that the universality is captured by a system-size independent relation $\tilde{V}    = f (\tau , a_{Z,\delta})$, where $\tau = (t - t_{\rm min})$ for the scaled variance $\tilde{V} = N^{-\nu} \mathrm{Var}[\mathcal{O}^-]$. We also define the scaled layer spacing $a_{Z,\delta} = L^{-\delta} a_Z$, where the exponent $\delta$ determines how $a_Z$ should be scaled to compensate for changes of the layer length scale $L=N^{1/d}$. We then use basic scaling theory \cite{Altland2010} to arrive at the scaling ansatz
\begin{equation}
\mathrm{Var}\left[\mathcal{O}^-\right]a_Z^{-d_{V}}N^{d_V\delta /d-\nu}=f\big [ (t-t_{\rm min}) a_Z^{-d_{\tau}} N^{\delta d_{\tau}/d} \big ]\label{eq:fullscale} 
\end{equation}
 The new exponents $d_V$ and $d_{\tau}$ characterize the divergence of $\tilde{V}$ and $t$, respectively, with $a_{Z,\delta}$. To demonstrate the efficacy of the ansatz let us first consider the simpler case where the system size $N$ is fixed. This allows $N$ to be absorbed into the scaling function, now $f_N$, and the ansatz reduces to
$
   \mathrm{Var}[\mathcal{O}^-] a_Z^{-d_{V}}    = f_N [ (t - t_{\rm min})a_Z^{-d_{\tau}} ]. 
$ 

To demonstrate this scaling, Fig.~\ref{fig:timecollapse}(a) shows  $  \mathrm{Var}[\mathcal{O}^-]a_Z^{-d_{V}} $ vs $(t - t_{\rm min})a_Z^{-d_{\tau}}$ for a variety of $a_Z$ values. The data corresponds to the case of $\alpha = 1.5$ with $d=1$, for which we can reach large enough system sizes to simulate  an order of magnitude of $a_Z$ values in the partially collective phase \cite{Footnote}. While the raw data varies substantially with $a_Z$ (inset), the scaled data collapse to a single  curve $f_N$ up until $t \sim t_{\rm min} $, after which point additional interaction effects are anticipated to become substantial. 
The inset of  Fig. \ref{fig:timecollapse}(b) shows these $f_N$ for a variety of $N$ (colors). In the main figure we demonstrate the full collapse of all data across $t$, $a_Z$ and $N$ by scaling both axes by the $N$-dependent terms in Eq. \ref{eq:fullscale}. %
 The scaling collapse is also effective for other $\alpha$ values in $1d$, and for accessible system sizes in $2d$ (SM~\cite{supplemental})

Finally, we clarify how the exponents relate to the scaling of variance minima considered in Fig. \ref{fig:fig2}. In this case $t = t_{\rm min}$, which reduces the rhs of Eq. \ref{eq:fullscale} to $f(0)$. Re-arranging the remaining terms, we obtain
\begin{equation}
\mathrm{Var}[\mathcal{O}^-]_{\rm min}  N^{-p}    =   y \big( a_{Z} / L \big), \label{eq:minscale}
\end{equation}
where $y(x) = f(0) x^{d_V}$ and $p = d_V(1-\delta)/d  + \nu$. This reveals that the scaling function $x^{d_V}$ shown in Fig. \ref{fig:fig2} itself contains universal information. Due to the simple form of $y(x),$ powers of $N$ can be moved from the lhs of the Eq. \ref{eq:minscale} to the right. Thus, the aspect ratio $a_Z/L$ characterizes the critical point itself, as shown in Fig. \ref{fig:fig2}, in the partially collective phase $\delta$ captures the relevant scaling of $a_Z$ with system size. Intuitively, if $\alpha < d$ such that the interactions are genuinely long-ranged, then $\delta > 1$ ensures that as the system size increases the range of $a_Z$ values below the critical value $(a_Z/L)_* = {\rm constant}$ reduces, i.e. the system becomes more collective for larger $L$ and it becomes   difficult to observe the partially collective phase.

 {\it Universality of the partially collective phase---}%
We have checked that the scaling exponents for square, triangular and hexagonal lattices in $2d$ for $\alpha = 2$ are consistent within error \cite{supplemental}, strongly indicating universality independent of microscopic details. %
However, we note that the specific spin-structure and symmetries of the interactions are critical for the observed phenomenology. Indeed, a fully symmetric XXX model only shows $N^{1/2}$ scaling in the fully collective phase \cite{PhysRevA.109.L061304}, worse than the partially collective phase observed for anisotropic spin interactions here, necessarily changing the character of both phases. %
Intuitively, the in-layer $SU(2)$ symmetry is crucial to gap-protect the collective manifold  (and suppress imaginary finite momentum modes in the Bogoliubov picture), while interlayer ZZ interactions suppress the creation of entangled spin-flips from the initial state. %
Breaking the global $U(1)$ symmetry (recognized to be critical for one-axis twisting squeezing dynamics \cite{Roscilde2022,Roscilde2023}), here suppresses the scaling from $1/N$ to $1/\sqrt{N}$ \cite{supplemental}, again changing the character of both phases.

{\it Experimental considerations---}%
Our predictions can be readily tested in state-of-the art experiments; 
bi-layer geometries can be realized for polar molecules \cite{Tobias2022}, magnetic atoms \cite{Du2024}, Rydberg atoms in reconfigurable tweezer arrays \cite{PhysRevLett.130.180601,Barredo2018,Bluvstein2023}, and trapped ion arrays \cite{Hawaldar2024}. 
 The necessary Floquet engineering of interactions has been demonstrated in all of these platforms \cite{Lukin_2020_Robust,Geier2021,Scholl2022,Christakis2023,Miller2024,PRXQuantum.4.010334}, and only requires the capability to selectively manipulate ensembles (here layers) of spins. %
Direct observation of the reduced variance would place stringent requirements on the measurement resolution, requiring single particle detection in the fully collective regime. 
This has been demonstrated in some platforms \cite{PhysRevLett.109.133603,PhysRevLett.111.253001} but is not yet widely available. %
Approaches to indirectly measure the reduced variance may relax that requirement \cite{wu2025spinsqueezingensemblenitrogenvacancy}, while time-reversal protocols \cite{PhysRevResearch.6.033197} allow using the enhanced sensitivity without single-particle detection efficiency \cite{PhysRevLett.116.053601}.

{\it Outlook---}%
Our work demonstrates nonequilibrium critical scaling in the two-mode squeezing dynamics of powerlaw interacting spin ladders and bi-layers. %
We find two dynamical squeezing phases, a fully collective phase and a partially collective phase, distinguished by the scaling of the achievable minimal variance. %
Crucially, we develop a scaling ansatz capturing the full finite time dynamics in the partially collective phase, which also determines the scaling of the minimal variance. This establishes these regimes as true dynamical phases, and the variance scaling exponent as an indicator of the phase transition. We further demonstrate independence of the critical exponents on lattice geometry, highly indicative of universality. Unlike prior work, the scalable squeezing we observe is not due to finite-temperature order. Given our initial state is an infinite temperature state of the inter-layer interactions, and the (staggered) magnetisation dynamically decays during the squeezing dynamics, scalable squeezing here appears as a genuine non-equilibrium phenomenon.

In addition, our work establishes scalable two-mode squeezing with powerlaw interactions (also at fixed layer-separation, Fig.~S10 \cite{supplemental}). This resolves the outstanding fundamental question of achieving scalable squeezing beyond the one-axis twisting paradigm in long-range \cite{Block_2024} or quasi-long range \cite{Roscilde2024} ordered phases. %
Our findings also suggest engineering the Bogoliubov excitation spectrum as a powerful tool for controlling squeezing dynamics. %

Finally, it may be interesting to connect the critical scaling found here with the non-equilibrium spatio-temporal scaling and universality class identified recently in short-range 2D Heisenberg models \cite{Rodriguez_Nieva_2022}.

 \begin{acknowledgments}
\noindent{\textit{Acknowledgements:}
The computing for this project was performed at the High Performance Computing Center at Oklahoma State University supported in part through the National Science Foundation grant OAC-1531128.
} 

\end{acknowledgments}

\nocite{Dresselhaus2022,Houdayer2004}

\bibliography{TMS_new}{}

\cleardoublepage

\end{document}


\title{Supplemental Material for ``Nonequilibrium Critical Scaling of a Squeezing Phase Transition"}

\author{Arman Duha}
\affiliation{Department of Physics, Oklahoma State University, Stillwater, Oklahoma 74078, USA}

\author{Samuel E. Begg}
\affiliation{Department of Physics, Oklahoma State University, Stillwater, Oklahoma 74078, USA}

\author{Thomas Bilitewski}
\affiliation{Department of Physics, Oklahoma State University, Stillwater, Oklahoma 74078, USA}

\date{\today}

\maketitle

The Supplementary Material contains additional details on symmetry-breaking perturbations, the scaling for other lattice geometries, Bogoliubov analysis, layer polarization and collectiveness, metrological utility, the optimization procedure used to estimate critical exponents, scaling analysis of the variance minima at fixed layer spacing, and extended data of the scaling collapse for power-law interaction exponents $\alpha$ and dimensions $d$.

\section{Symmetry-Breaking Perturbations}
\begin{figure}[t]
\includegraphics[width=\columnwidth]{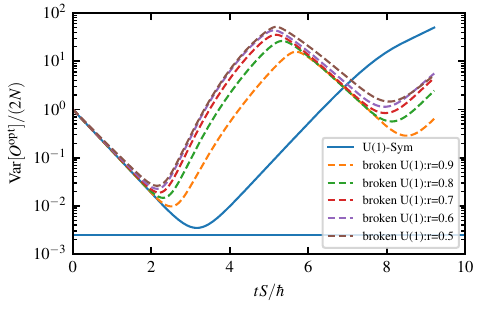}
\includegraphics[width=\columnwidth]{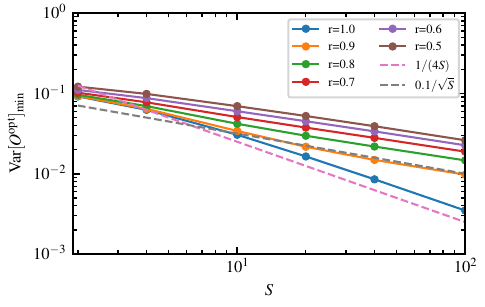}
\caption{Broken U(1) Symmetry. Exact dynamics of the infinite range-model ($\alpha=0$) for inter-layer XY anisotropic interactions measured by the ratio $r$ of XX to YY interactions. Top panel time-evolution of the variance of the optimal observable for $S=N/2=100$ for a range of $r$ (legend). Horizontal line is the Heisenberg limit. Bottom panel scaling of the minimal variance of optimal observable with system-size $S$ for different anisotropies $r$ ($r=1$ being isotropic).
\label{response_fig:broken_U1}}
\end{figure} 
In this section we consider symmetry-breaking perturbations to the model considered in the main text.

We demonstrate that the global U(1) symmetry of the Hamiltonian Eq.~(1) is essential for the observed phases and phenomenology. For this it is sufficient to consider the infinite range ($\alpha=0$) case, in which we can perform exact time-dynamics at moderate system sizes (even without U(1) symmetry). To this end we consider symmetry broken Hamiltonians of the form $\sqrt{r} S_A^X S_B^X + 1/\sqrt{r} S_A^Y S_B^Y$, where $S_A^{\alpha}$ are the total layer-spins in layer $\alpha=A,B$, and $r$ measures the anisotropy as the ratio of the strength of XX to YY interactions between layers, and we dropped the intra-layer Heisenberg interactions, which only contribute a constant in the fully collective limit. %

We show in the top panel of Fig.~\ref{response_fig:broken_U1} the time-evolution of the variance of the optimal observable for these hamiltonians at fixed system size $S=N/2$ for different anisotropy ratios $r$ (left panel), demonstrating how anisotropy negatively impacts the achievable squeezing. %
The optimal observable here is defined as the eigenvector with minimal eigenvalue of the full co-variance matrix of layer-observables, which for $r=1$ is $O^{-}$, but for finite anisotropy generically differs, or alternatively the plot may be understood as the minimal eigenvalue of the full co-variance matrix. %
We show the scaling of the minimal variance with system size $S$ in the bottom panel, demonstrating that finite anisotropy reduces the achievable scaling from $1/S\sim 1/N$ to $1/\sqrt{S} \sim 1/\sqrt{N}$.

\begin{figure*}[hptb]
\includegraphics[width=\textwidth]{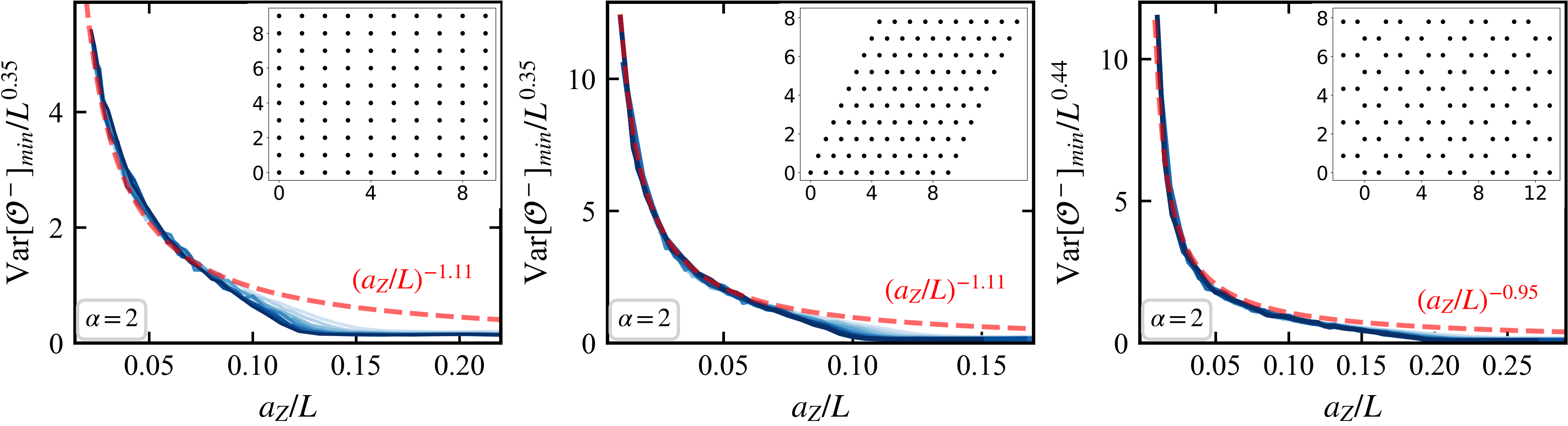}
\caption{Universal scaling collapse of minimal squeezing for square, triangular, and hexagonal lattices. Rescaled minimal variance $\text{Var}[\mathcal{O}^-]_{\rm min}/L^{p \times d}$ versus aspect ratio $a_Z/L$ for $L$ = 30 to 100 (opacity increases with system size), demonstrating a collapse onto a single curve. The red dashed lines $(a_Z/L)^{d_V}$ show fit to the scaling function with $d_V$ from Table.~\ref{tab:exponents_geometry} 
}
\label{fig:diff_lattices}
\end{figure*}

\begin{figure*}
\includegraphics[width=\textwidth]{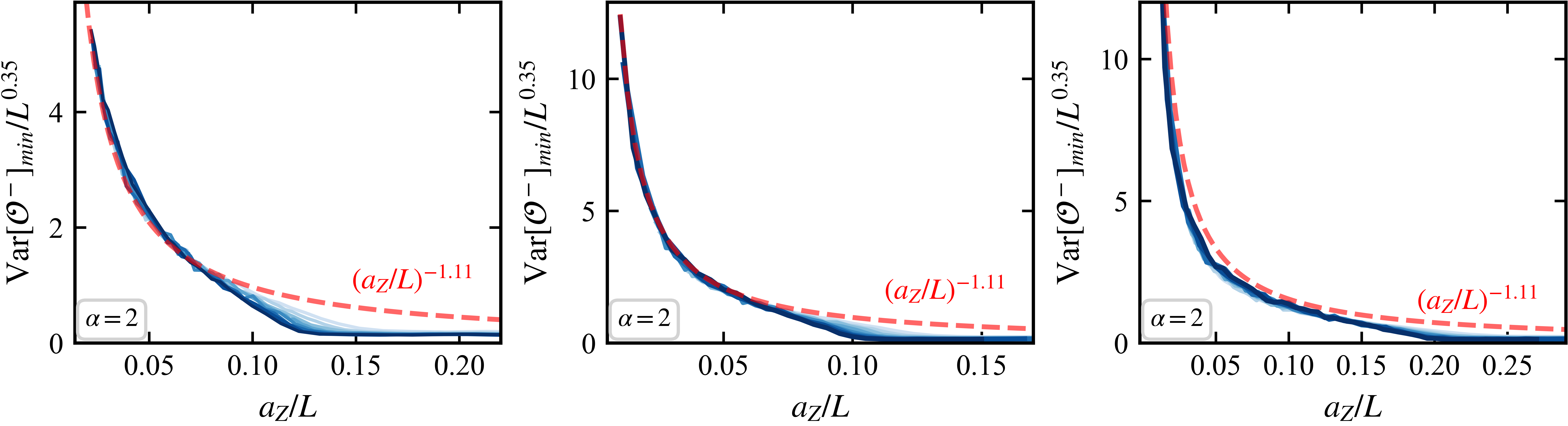}
\caption{Universal scaling collapse of minimal squeezing for square, triangular, and hexagonal lattices using the exponents obtained for the case of the square lattice. Rescaled minimal variance $\text{Var}[\mathcal{O}^-]_{\rm min}/L^{p \times d}$ versus aspect ratio $a_Z/L$ for $L$ = 30 to 100 (opacity increases with system size), demonstrating a collapse onto a single curve. The red dashed lines $(a_Z/L)^{d_V}$ show fit to the scaling function using the $d_V$ values obtained for the case of the square lattice.}
\label{fig:diff_lattices_same_exp}
\end{figure*}

\section{Universality across different lattices}
We now provide evidence that the same set of exponents can describe the partially collective phase with different lattice geometries. Similar to the square lattice case, the lattice spacing for the triangular and hexagonal lattices is set to $a_{\text{lat}} = 1$ (see Fig. \ref{fig:diff_lattices}). For simplicity, we focus on the case of $\alpha = 2$ in $2d$. In Fig. \ref{fig:diff_lattices} we present data for $\text{Var}[\mathcal{O}^-]_{\rm min}/L^{p \times d}$ vs $a_Z/L$ for square, triangular and hexagonal unit cells. The y-axes are scaled by the exponents listed in Table \ref{tab:pval_geometry}. Under these rescalings, the data collapses to a single curve, or scaling function, 
and it can be seen that the aspect ratio $a_Z/L$ controls the transition. Crucially, the rescaling exponents for the three lattice types agree with each other within standard error. %
We explicitly demonstrate in Fig. \ref{fig:diff_lattices_same_exp} that the same set of scaling exponents allows to collapse all three lattices. %
This is strong evidence for the universal nature of the transition
as the exponents remain invariant under this change of geometry and associated change in the microscopic interactions. %
Similarly, the dashed line $(a_Z /L)^{d_V}$ is an excellent fit to the scaling function and the exponents $d_V$ for all three lattice types agree within error (see Table \ref{tab:exponents_geometry}).

\begin{figure*}[p]
\begin{center}
\subfloat[Triangular, $N = 10000$]{
    \includegraphics[width = 7cm]{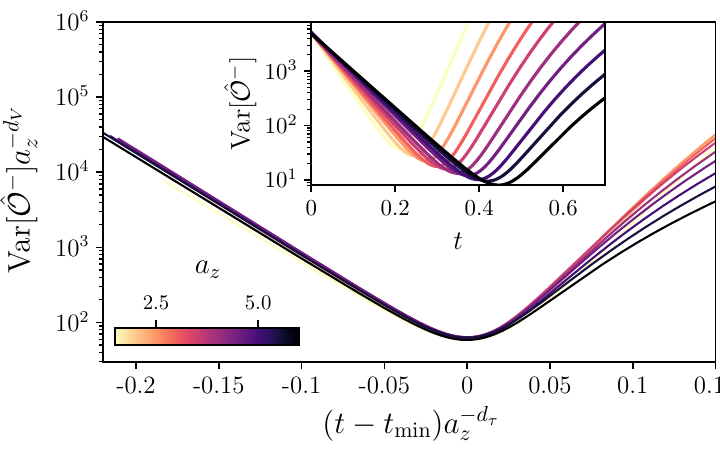} }
\subfloat[Triangular]{
    \includegraphics[width = 7cm]{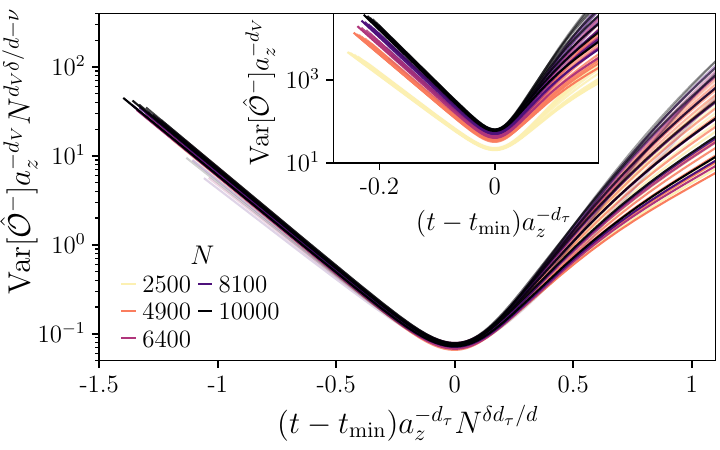} }
    
    \subfloat[Hexagonal, $N = 10000$]{
    \includegraphics[width = 7cm]{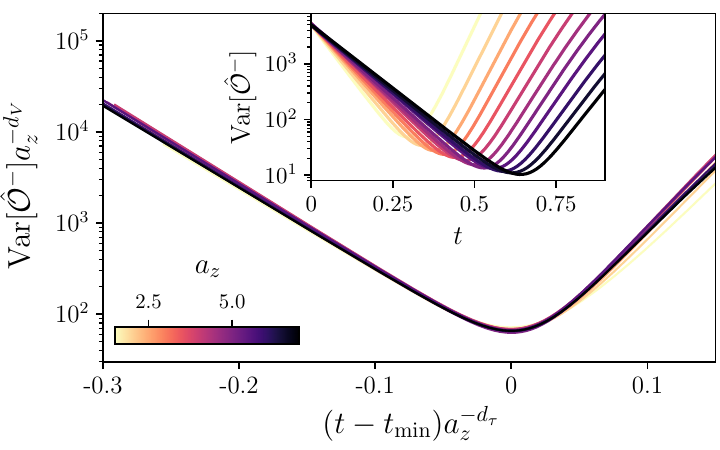} }
\subfloat[Hexagonal]{
    \includegraphics[width = 7cm]{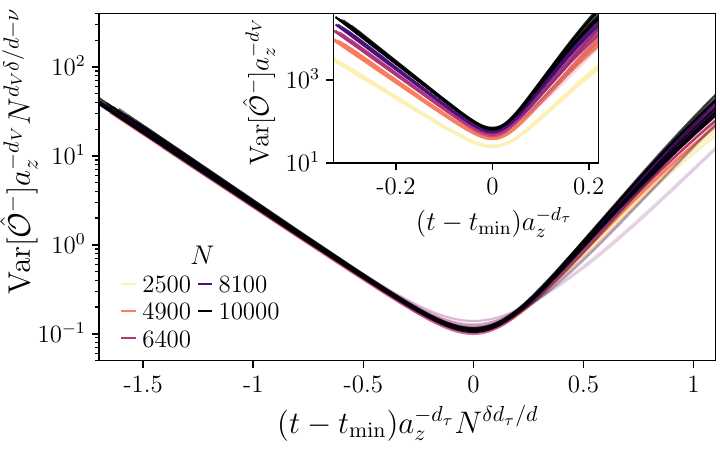} }

\subfloat[Multiple lattice types, $N=10000$]{
    \includegraphics[width = 7cm]{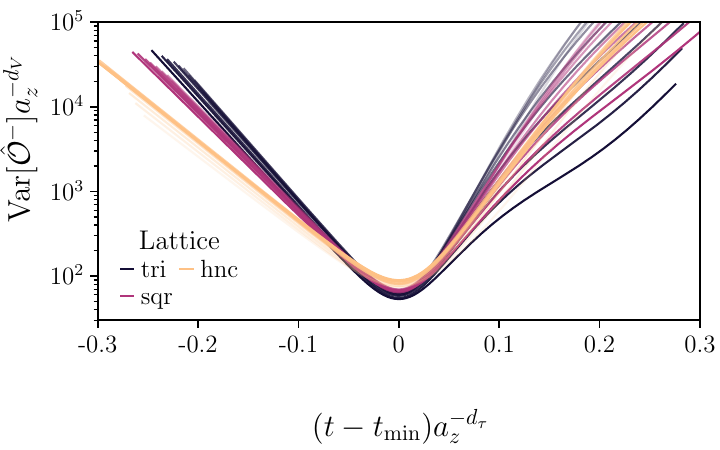} }
\end{center}
    \caption{Raw data and scaling collapses for $\alpha=2$ in $2d$, for the triangular lattice  [(a) and (b)] and the hexagonal lattice  [(c) and (d)]. Plot (e) plots the scaling collapse for the square (sqr), triangular (tri) and hexagonal (hnc) lattices on the same figure. Lines for smaller $a_Z$ values are more faded in (b), (d) and (e). These exponents are the average of the exponents for the three individual lattice geometries, provided in Table \ref{tab:exponents_geometry}.}
    \label{fig:SI_lattice_timescaling}
\end{figure*}

In Fig. \ref{fig:SI_lattice_timescaling} we present results for the full scaling collapse as a function of time $t$, layer spacing $a_Z$, and system size $N$. The triangular and hexagonal lattice geometries show a similar collapse to the square lattice case discussed in the main text. Again, all exponents for the three lattice types are within standard error, and are therefore compatible with universality; see Table \ref{tab:exponents_geometry}.  In Fig. \ref{fig:SI_lattice_timescaling}(e) we present data for all three unit cells for a single set of exponents. These exponents are the average of the exponents for the three individual lattice geometries, provided in Table \ref{tab:exponents_geometry}. %
Interestingly, while the scaling collapse works across lattices, the scaling functions themselves show notable discrepancies, particularly at early times. 

\section{Bogoliubov Analysis}
In this section we analyze the dynamics within a non-interacting approximation, via the Holstein-Primakoff transformation to bosons. This follows the general construction of Refs. \cite{Bilitewski2023} and \cite{Dominguez2024}.  For the A layer, the mapping to bosons $\hat{a}_{\bm j}$ is
\begin{align}
   &  \hat{s}^+_{A \bm{j} } = \sqrt{1- \hat{a}_{\bm{j} } ^{\dagger} \hat{a}_{\bm{j} } }~ \hat{a}_{\bm{j} }   \\ &
    \hat{s}^-_{A\bm{j} }    = \hat{a}^{\dagger}_{\bm{j} } \sqrt{1- \hat{a}_{\bm{j} } ^{\dagger} \hat{a}_{ \bm{j} } }    \\ &
      \hat{s}^z_{A\bm{j} }   = \frac{1}{2} - \hat{a}^{\dagger}_{\bm{j} } \hat{a}_{\bm{j} }   ,
\end{align}
while for layer B we introduce bosons $\hat{b}_{\bm j}$ 
\begin{align}
   &  \hat{s}^-_{B\bm{j} } = \sqrt{1- \hat{b}_{\bm{j} } ^{\dagger} \hat{b}_{ \bm{j} } }~ \hat{b}_{j}   \\ &
    \hat{s}^+_{B \bm{j} }    = \hat{b}^{\dagger}_{ \bm{j} } \sqrt{1- \hat{b}_{ \bm{j} } ^{\dagger} \hat{b}_{\bm{j} } }    \\ &
      \hat{s}^z_{B\bm{j} }   = -\frac{1}{2} + \hat{b}^{\dagger}_{ \bm{j} } \hat{b}_{ \bm{j} }   ,
\end{align}
where $[\hat{a}_{\bm i},\hat{a}^{\dagger}_{\bm j}] = [\hat{b}_{\bm i},\hat{b}^{\dagger}_{\bm j}] = \delta_{\bm{i} \bm{j}}$ 
and all other commutators vanish. Note that in this representation the vacuum state is $\ket{\uparrow}$ ($\ket{\downarrow}$) for the A layer (B layer), which are chosen due to align with our choice of initial state.
We expand the square root to first order in the density using $\sqrt{1 - \hat{n}_{A/B} } = 1 - \frac{1}{2}\hat{n}_{A/B} + \mathcal{O}(n_{A/B}^2)$. These are substituted into the Hamiltonian in Eq. (\ref{eq:model}) of the main text. After normal ordering the result and truncating to $\mathcal{O}(a^2)$ we obtain 
\begin{align}
& \hat{H} =  \frac{1}{2}\sum_{\bm{l}\neq\bm{j}} V_{\bm{l}\bm{j}}\Big[ 
\hat{a}^{\dagger}_{\bm{l}} \hat{a}_{\bm{j}} + \hat{a}^{\dagger}_{\bm{j}} \hat{a}_{\bm{l}}   + \hat{b}^{\dagger}_{\bm{l}} \hat{b}_{\bm{j}}  + \hat{b}^{\dagger}_{\bm{j}} \hat{b}_{\bm{l}} \nonumber \\ &  - 2 ( \hat{n}_{A \bm{j}} +  \hat{n}_{B \bm{j}} )    \Big] + \sum_{\bm{l},\bm{j}} V_{\bm{l}\bm{j}}  \Big( \hat{a}_{\bm{l}}\hat{b}_{\bm{j}} + \hat{a}^{\dagger}_{\bm{l}} \hat{b}^{\dagger}_{\bm{j}}      \Big). \label{eq:model_bosons} 
\end{align} 
The interaction strength depends on only the distance between the spins, i.e. $V_{\bm{l} \bm{j}} = V_{\bm{l}-\bm{j}}$. 
For the in-layer interactions we therefore define
the Fourier transform as $
     V^{||}_{\bm{k}}   =  \frac{1}{L^{d/2}} \sum_{\bm{j} } e^{-i \bm{j} \bm{k}}  V_{\bm{j}}$, where we have set the lattice spacing $a = 1$ and $\bm{k}$ takes values in the two-dimensional Brillioun zone. Similarly, for interlayer interactions 
$
   V^{\perp}_{\bm{k}}    =  \frac{1}{L^{d/2}} \sum_{\bm{j}\in {A - B}  } e^{-i\bm{j}\bm{k}}  V_{\bm{j}}$, where the summation is conducted over all vectors connecting a site in layer A to one in layer B. 
      Fourier transforms of the operators $\hat{a}_{\bm{j}}$ and $\hat{b}_{\bm{j}}$ also follow the standard definitions. Moving to momentum space,  Eq. (\ref{eq:model_bosons}) becomes \cite{Bilitewski2023,Dominguez2024}
\begin{align}
    \hat{H}  = &  \sum_{\bm{k}} \Big[ \varepsilon_{\bm {k}} (
\hat{a}^{\dagger}_{\bm{k}}    \hat{a}_{\bm{k}} + \hat{b}^{\dagger}_{\bm{k}}    \hat{b}_{\bm{k}}) +  \Omega_{\bm{k}} \hat{a}_{-\bm{k}}    \hat{b}_{\bm{k}}  + \Omega_{\bm{k}} ^* \hat{a}^{\dagger}_{-\bm{k}} \hat{b}^{\dagger}_{\bm{k}}  \Big]  ,
\end{align} 
where $
    \Omega_{\bm{k}} =  \frac{1}{L^{d/2}}V^{\perp}_{\bm{k}}$  and $ \varepsilon_{\bm{k}}  = \frac{1}{L^{d/2}} (V^{||}_{\bm{k}}  -V_0^{||} ) 
$.
This can be diagonalized via a Bogoliubov transformation, leading to 
\begin{align}
    \hat{H} & = \sum_k   \sqrt{\varepsilon_{\bm{k}}^2 -|\Omega_{\bm{k}}|^2} \Big(\hat{\beta}^{\dagger}_{\bm{k}}  \hat{\beta}_{\bm{k}} + \hat{\gamma}^{\dagger}_{\bm{k}}\hat{\gamma}_{\bm{k}} \Big),
\end{align}
where the new operators $\hat{\beta}_{\bm{k}}$ and $\hat{\gamma}_{\bm{k}} $  obey the bosonic commutation relations $[\hat{\gamma}_{\bm{k}}, \hat{\gamma}^{\dagger}_{\bm{k'}} ] = [\hat{\beta}_{\bm{k}}, \hat{\beta}^{\dagger}_{\bm{k'}} ] = \delta_{\bm{k}\bm{k'}}$ with all other commutators vanishing. We then solve the Heisenberg equations 
$ \partial_t  \hat{\beta}_{\bm{k}}  = i [\hat{H}_{\bm{k}},  \hat{\beta}_{\bm{k}}] $  and $ \partial_t  \hat{\gamma}_{\bm{k}}  = i [\hat{H}_{\bm{k}},  \hat{\gamma}_{\bm{k}}] $
to give 
\begin{align}
  \hat{\beta}_{\bm{k}} (t) = e^{-i \sqrt{\varepsilon_{\bm{k}}^2 -|\Omega_{\bm{k}}|^2} t}\hat{\beta}_{\bm{k}} (0), \label{eq:beta}\\ 
   \hat{\gamma}_{\bm{k}} (t)  = e^{-i \sqrt{\varepsilon_{\bm{k}}^2 -|\Omega_{\bm{k}}|^2} t}\hat{\gamma}_{\bm{k}} (0). \label{eq:gamma} 
\end{align}
which clearly illustrates that unstable modes with $\xi_{\bm{k}} = \varepsilon_{\bm{k}}^2 -|\Omega_{\bm{k}}|^2 < 0$ will grow exponentially. Since this
generates the dominant contribution to the squeezing we now restrict our attention to the evolution of these unstable modes. In terms of the original bosonic variables, Eq. (\ref{eq:beta}) and (\ref{eq:gamma}) (for the unstable modes) is  
\begin{align}
  \hat{a}^{\dagger}_{\bm{k}}(t)   = &
 \hat{a}^{\dagger}_{\bm{k}}(0) \Big( i\sinh (|\xi_{\bm{k}}|t) \frac{\varepsilon_{\bm{k}}}{|\xi_{\bm{k}}|} + \cosh (|\xi_{\bm{k}}|t) \Big)     \nonumber \\ &  + i  \hat{b}_{-\bm{k}}(0) \sinh (|\xi_{\bm{k}}|t)  \frac{\Omega^*_{\bm{k}}}{|\xi_{\bm{k}}|} .  \label{eq:a_ev}  \\
      \hat{b}_{-\bm{k}}(t)  =   & 
     \hat{b}_{-\bm{k}} (0) \Big(-i \sinh (|\xi_{\bm{k}}|t) \frac{\varepsilon_{\bm{k}}}{|\xi_{\bm{k}}|} + \cosh (|E_k|t)   \Big) 
  \nonumber  \\ &   - i \hat{a}^{\dagger}_{\bm{k}}(0)  \sinh (|\xi_{\bm{k}}| t)
     \frac{\Omega_{\bm{k}}}{|\xi_{\bm{k}}|}  , 
  \label{eq:b_ev}   
\end{align}
Using the fact that the initial state is a vacuum state in the bosonic framework, simple manipulations give 

\begin{align}   
 \langle \hat{a}^{\dagger}_{\bm{k}}\hat{a}_{\bm{k}} \rangle(t) = \langle \hat{b}^{\dagger}_{\bm{k}}\hat{b}_{\bm{k}}\rangle  (t)
 & =
\sinh^2 (|\xi_{\bm{k}}|t)  \frac{|\Omega_{\bm{k}}|^2}{|\xi_{\bm{k}}|^2}  \\ & \propto e^{2|\xi_{\bm{k}}|t}  \frac{|\Omega_{\bm{k}}|^2}{|\xi_{\bm{k}}|^2}   \nonumber ,
\end{align} 
where the last line indicates the exponential growth that occurs after an initial transient period.
We now consider the squeezed quadratures $ \mathcal{O}^- = \hat{S}_B^x - \hat{S}_A^y$ and $\hat{S}_B^y + \hat{S}_A^x $, and anti-squeezed quadratures $ \mathcal{O}^+ = \hat{S}_B^x + \hat{S}_A^y$ and $\hat{S}_B^y - \hat{S}_A^x $.
Since we are working at quadratic order we can replace the collective spin operators with the bosonic operators, i.e. $S^{+}_A \rightarrow \hat{a}_{\bm{k}=0} $ and $S^{+}_B \rightarrow \hat{b}^{\dagger}_{\bm{k}=0} $. Using Eqs. (\ref{eq:a_ev}) and (\ref{eq:b_ev}), and after some manipulation, we derive the variances 
\begin{align}
 {\rm{Var}}\Big[ \mathcal{O}^{\pm} (t) \Big] =  \frac{1}{4}(A_0(t) + B_0(t))^{\pm 2}  ,
\end{align}
where
\begin{align}
& A_0 = \sqrt{\sinh^2 (|\xi_0|t) \frac{\varepsilon^2_{0}}{|\xi_0|^2} + \cosh^2(|\xi_0| t) }  \\ &
B_0 = \sinh (|\xi_0| t)
     \frac{|\Omega_{0}|}{|\xi_0|}  ,
\end{align}
and we have used the abbreviation $E_{\bm{k}=0} = \xi_0$.
Beyond short times, we can substitute $\sinh{|\xi_0| t} \approx \frac{1}{2} e^{|\xi_0| t}$ to obtain
\begin{align}
 & {\rm{Var}}\Big[ \mathcal{O}^{\pm}(t) \Big]  
\approx   \frac{e^{\pm 2 |\xi_0|t}}{4}\Big( \frac{1}{2}\sqrt{1 +  \frac{\varepsilon^2_{0}}{|\xi_0|^2}}  + 
     \frac{|\Omega_{0}|}{|\xi_0|}  \Big)^{\pm 2}. 
\end{align}

\subsection{Analysis of Unstable Modes}
Based on the above Bogoliubov analysis the $k=0$ mode will always be unstable resulting in exponential growth of excitations in the collective manifold, and the associated exponential squeezing of the corresponding quadrature. %
However, finite $k$ modes may also be unstable depending on model parameters. We therefore now analyse how the existence of these unstable modes varies as a function of system parameters.

\begin{figure}[t]
\includegraphics[width=\columnwidth]{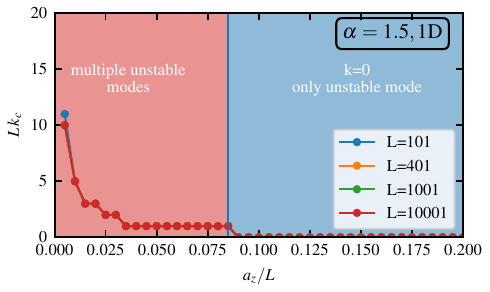}
\caption{Largest unstable mode $k_c$ scaled by linear system size $L$, as a function of the aspect ratio $a_Z/L$ for a range of system sizes $L$ (legend) in 1D for $\alpha=1.5$. 
\label{fig:fig_Bog_1}}
\end{figure} 
In Fig.~\ref{fig:fig_Bog_1} we show the largest momentum $k_c$ (scaled by linear system size $L$), which has imaginary Bogoliubov eigenenergy, as a function of the aspect ratio $a_Z/L$. 
We observe a well-defined transition point $(a_Z/L)_*$ (independent of system size), above which only the $k=0$ mode has imaginary eigen-energy, and below which there are multiple unstable modes with imaginary eigenenergies.

\subsection{Phase Transition}
\begin{figure}[hptb]
\includegraphics[width=\columnwidth]{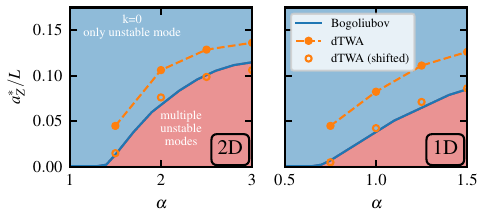}
\caption{Comparison of dTWA and Bogoliubov. Critical aspect ratio $(a_Z/L)_*$ (Bogoliubov, solid line) as a function of powerlaw exponent $\alpha$ separating the region, in which only the $k=0$ mode is unstable, and the region with multiple unstable modes. Compared with results from dTWA based on the scaling of the minimal variance (Fig 1, main text, and Fig.~\ref{fig:SI_MuVsaZL}).
\label{fig:fig_Bog_3}}
\end{figure} 
Based on the quadratic Bogoliubov theory, we may therefore distinguish two parameter regimes of the model. %
One in which only the $k=0$ mode is unstable, is predicted to grow exponentially, and generate entanglement in the form of two-mode squeezing, while all other momentum modes are stable and remain at low population. This would therefore correspond to the collective squeezing phase identified in the dTWA results. %
In contrast, in the regime with multiple unstable modes, the exponential growth of the finite k-modes (if it were to continue unabated) would rapidly reduce both the collective spin length and depolarise the system. %
Based on this intuition, the quadratic Bogoliubov theory would predict the existence of a fully collective regime, with a transition into a non-squeezing phase at a critical $a_Z/L$. 

We compare in Fig.~\ref{fig:fig_Bog_3} the phase diagram obtained from Bogoliubov theory, with the phase-boundary obtained from the dTWA simulations. While the transition point is not quantitatively captured in the Bogoliubov results (solid line), we observe good qualitative agreement in the shape of the transition line between the dTWA results (filled markers) and shifted dTWA results (empty markers).

This agreement indicates that indeed the absence of unstable Bogoliubov modes is related to the existence of the fully collective phase and the appearance of multiple unstable modes results into a transition out of the collective squeezing phase. 

\subsection{Partially Collective Phase}
These phase-diagrams indicate that the partially collective phase appears in a region with multiple unstable modes.

Naively, in the presence of multiple unstable modes one would not expect any scalable squeezing to persist based on the following argument. %
Using the predicted growth of the unstable modes $N_k \sim e^{\Gamma_k t}$ with
$\Gamma_k = 2{\rm Im}[\xi_{k}]$ 
(see Eq.~(S14)) and the fact that the most unstable mode is always $k=0$, with the second-most unstable mode being $k_1=1/L$, we may estimate the populations of these modes at the time of optimal squeezing, assuming scalable squeezing. %
We found that in the partially collective phase the minimal variance scales as $N^p$, which for two-mode squeezing is directly related to the number of generated pairs as $\mathrm{Var}_{\mathrm{min}} \sim 1/N_0$. %
Therefore, at the point of minimal variance $e^{-\Gamma_0 t_{\mathrm{min}}} = N^{-p}$ and consequently $t_{\mathrm{min}} = \log(N) p /\Gamma_0$. The number of particles with momenta $k_1$ that have been produced by this time is 
\begin{equation}
    N_1(t_{\mathrm{min}}) \sim e^{\Gamma_1 t_{\mathrm{min}}} = e^{\Gamma_1/\Gamma_0 \log(N) p} = N^{p\Gamma_1/\Gamma_0}.\label{eq:N1estimate}
\end{equation}
We now restrict our attention to $\alpha=3$ in 2D where the Bogoliubov excitation spectra is exactly solvable. In this case $\Gamma_0 = 1/a_Z$, 
$k_1 = 1/L$, and assuming a linear dependence for sufficiently small $k_1$, i.e. $\Gamma_1(k_1) \approx \Gamma_0 - \alpha k_1$, Eq. (\ref{eq:N1estimate}) becomes
\begin{equation}
    N_1(t_{\mathrm{min}}) \sim N^{p (1 - \alpha a_Z/L)}
\end{equation}
We find that based on the quadratic Bogoliubov theory we would expect a macroscopic population, scaling with a power of $N$ and large compared to order 1, in the finite momentum modes at the point of the observed optimal squeezing in the partially collective phase. In particular, the growth of unstable modes should be independent of $L$ at fixed $a_z/L$ (for sufficiently large $L$). This is clearly in conflict with the dTWA results which show that the system becomes more collective with increasing system size, and does not (significantly) depolarise until after the time of optimal squeezing (see Figs.~(\ref{fig:fig_SBzmSAz}-\ref{fig:fig_SL})).

Thus, interaction effects are 
critically important 
in order to stabilize the partially collective phase, and its existence is beyond a simple separation into a collective ``rotor" degree of freedom and non-interacting Bogoliubov excitations \cite{Roscilde2024,Roscilde2023}. This further provides strong evidence of the interacting nature of the partially collective phase found here, emerging from the interplay of multiple interacting modes, and  going beyond scenarios for scalable squeezing described previously.

\section{Layer polarization \& Collectiveness}
\begin{figure}[t]
\includegraphics[width=\columnwidth]{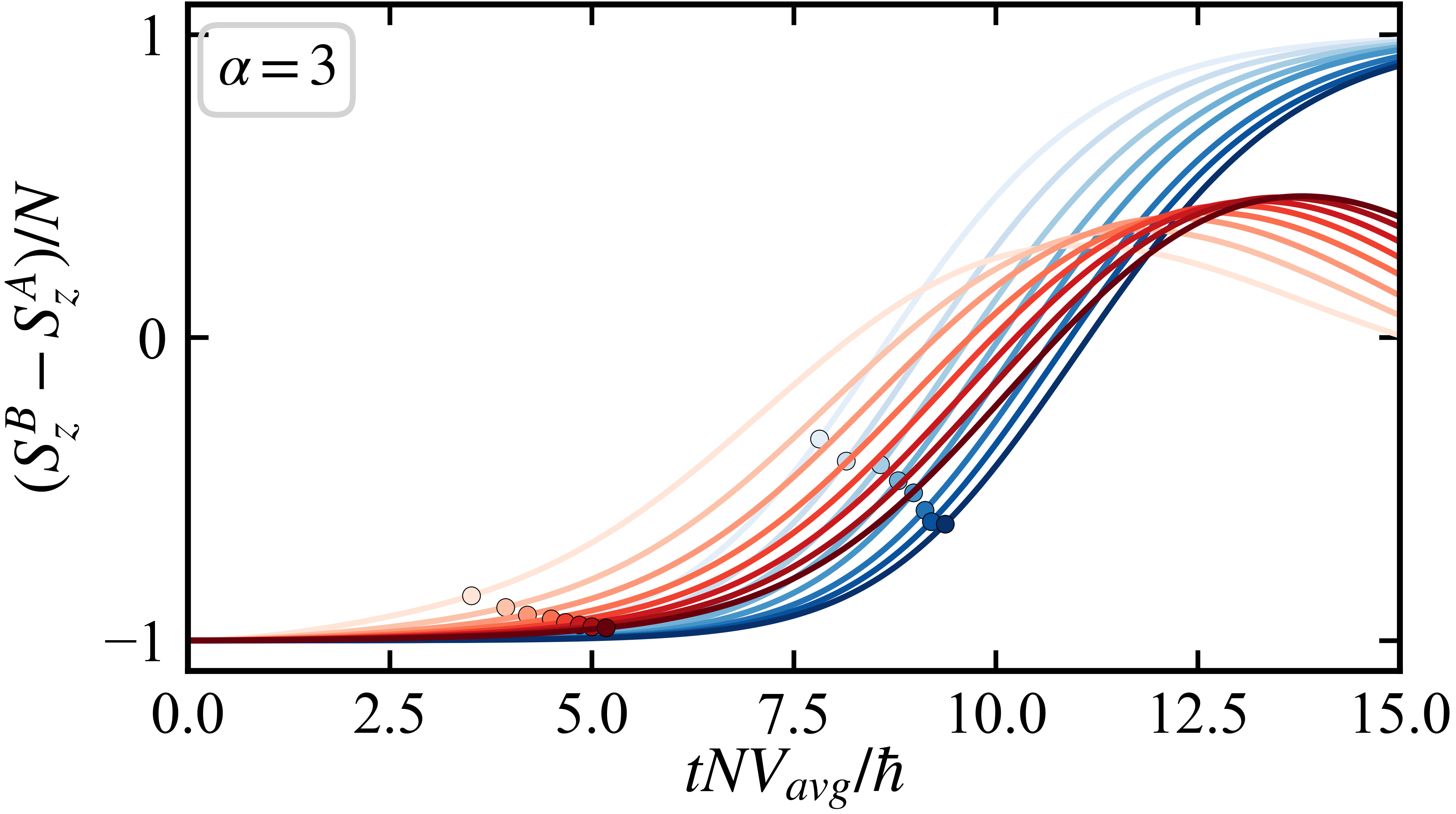}
\caption{Time evolution of $ (S^B_z - S^A_z)/N$ in the fully collective regime (blue, $aZ /L$ = 0.17) and partially collective
regime (red, $aZ /L$ = 0.04) for $\alpha$ = 3 in $2d$. The opacity increases with system size ($L$ = 30 to 100). The circles correspond to the time of optimal squeezing.
\label{fig:fig_SBzmSAz}}
\end{figure} 

In Fig.~\ref{fig:fig_SBzmSAz} we show the time-evolution of the layer polarization $\langle S_z^B-S_z^A\rangle$ for $\alpha=3$ in 2D for two values of $a_Z/L$ in the partially collective (red) and fully collective (blue) regimes for a range of system sizes $L$, i.e. for the same parameters as in Fig.~1(a) of the main text. %

We observe that the layer polarization is slightly reduced in the partially collective phase compared to the fully collective regime (which follows the two-mode squeezing prediction of exponential creation of pairs of spin flips). However, the decay is not substantial at the optimal squeezing time.

\begin{figure}[t]
\includegraphics[width=\columnwidth]{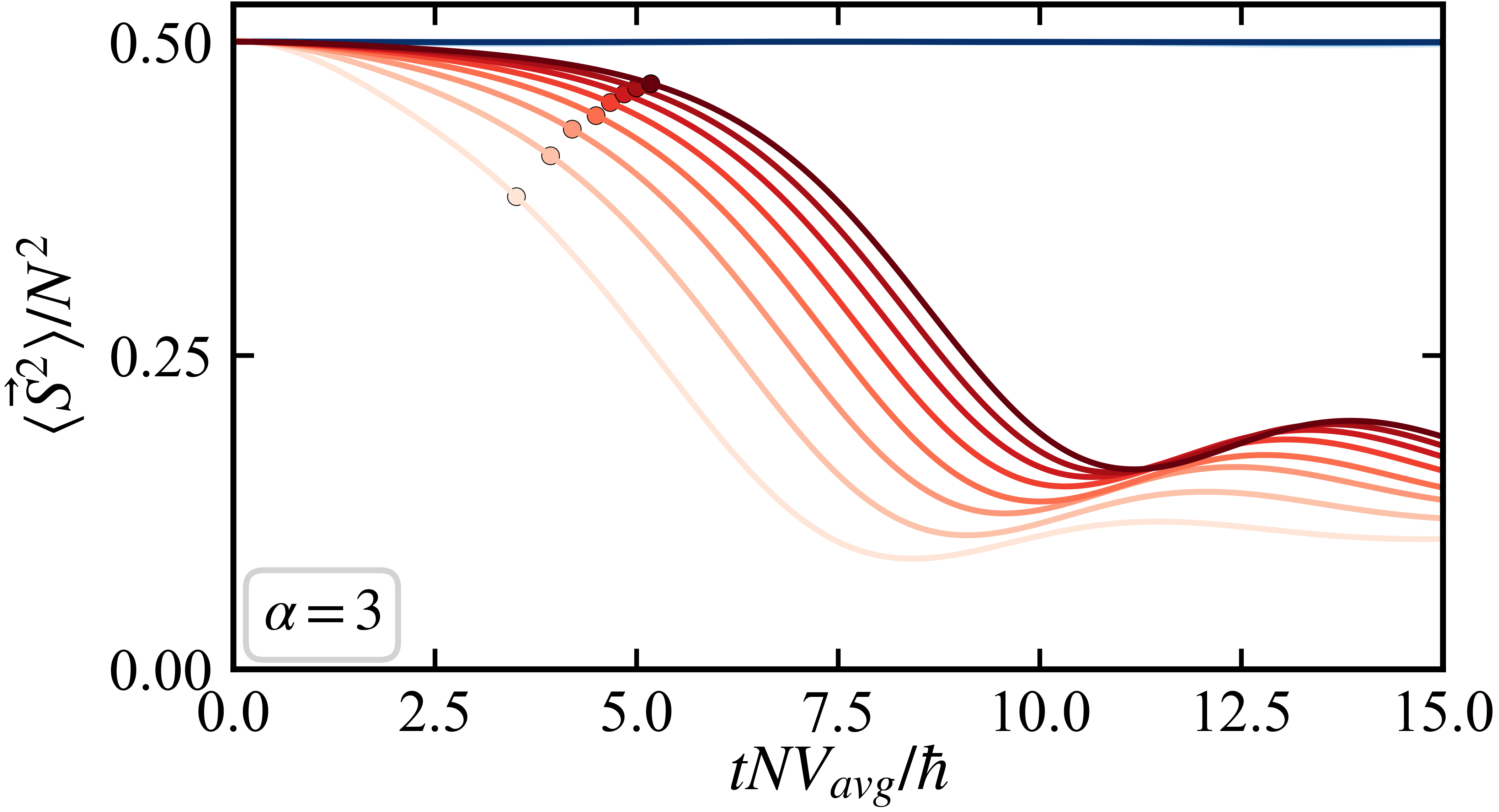}
\caption{Time evolution of length of collective layer spins $\langle \vec{S_A}^2 + \vec{S_B}^2 \rangle$, with $\hat{S}^{\mu}_{A (B)} = \sum_{\boldsymbol{i} \in A (B)} \hat{s}_{\boldsymbol{i}}^{\mu}$, in the fully collective regime (blue, $aZ /L$ = 0.17) and partially collective
regime (red, $aZ /L$ = 0.04) for $\alpha$ = 3 in $2d$. The opacity increases with system size ($L$ = 30 to 100). The circles correspond to the time of optimal squeezing. 
\label{fig:fig_SL}}
\end{figure} 
In Fig.~\ref{fig:fig_SL} we show the time-evolution of the length of the collective layer-spins for $\alpha=3$ in 2D for two values of $a_Z/L$ in the partially collective (red) and fully collective regime (blue) for a range of system sizes $L$, i.e. for the same parameters as in Fig.~1(a) of the main text. We observe that the layer spin length is slightly reduced in the partially collective phase, but scales towards becoming more collective at the time of optimal squeezing (as indicated by the open circles). In contrast, in the fully collective phase the layer spin length remains constant.

\section{Metrological utility and sensitivity}
Here, we briefly comment on the connection of the exponentially reduced variance of the squeezed quadratures to quantum enhanced sensitivity of a measurement based on standard arguments \cite{Sundar2023,PhysRevA.109.L061304}.

The squeezed variance directly results in an improved sensitivity of measuring a phase of rotation $\phi$ around $\hat{S}^z_A-\hat{S}^z_B$ \cite{Sundar2023} as
\begin{equation}
(\Delta \phi)^2 =\frac{(\Delta \mathcal{O})^2}{ (\langle \hat{S}_A^z - \hat{S}_B^z \rangle)^2}
\end{equation}
corresponding to an enhancement of $2 N (\Delta \phi)^2$ over using $2N$ unentangled particles.
While the variance decreases exponentially in time, at the same time the layer magnetisation decays as pairs of spin-flips are generated decreasing the sensitivity. 

In the fully collective regime using the two-mode squeezing prediction for the variance and number of excitations the optimal scaled sensitivity is $27/(8 N)$, retaining Heisenberg limited scaling \cite{PhysRevA.109.L061304}.

\begin{figure}[t]
\includegraphics[width=\columnwidth]{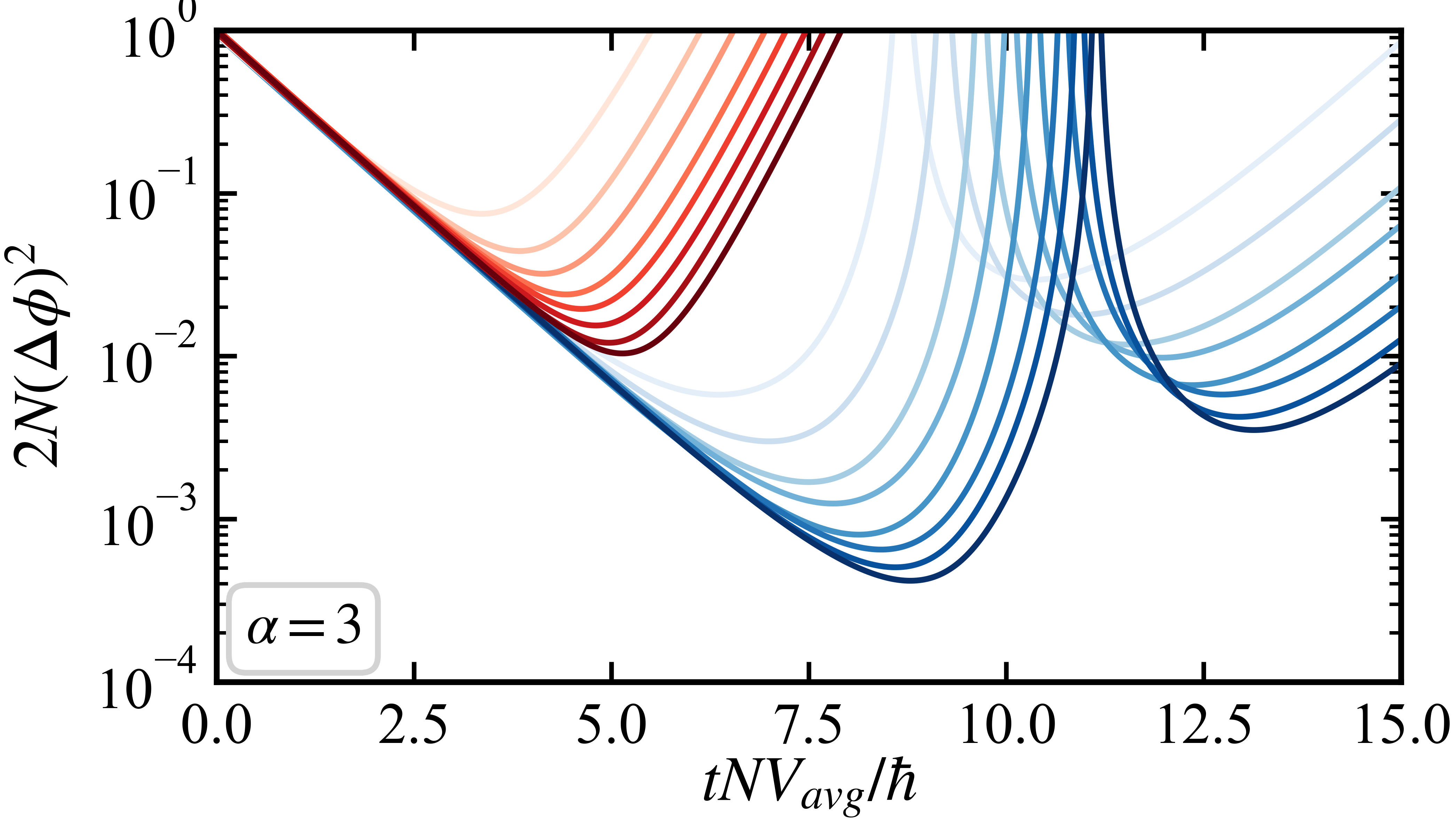}
\caption{Time evolution of scaled sensitivity in the fully collective regime (blue, $aZ /L$ = 0.17) and partially collective regime (red, $aZ /L$ = 0.04) for $\alpha$ = 3 in $2d$. The opacity increases with system size ($L$ = 30 to 100).
\label{fig:fig_sensitivity}}
\end{figure}

Since the decay of the layer polarization is sub-dominant for both the fully collective and partially collective regimes,  the scaling of the variance determines the resulting scaling of the sensitivity. For completeness, we explicitly show results for this sensitivity in Fig.~\ref{fig:fig_sensitivity}.

\section{Scaling of squeezing at fixed layer separation}
\begin{figure}[hptb]
\includegraphics[width=\columnwidth]{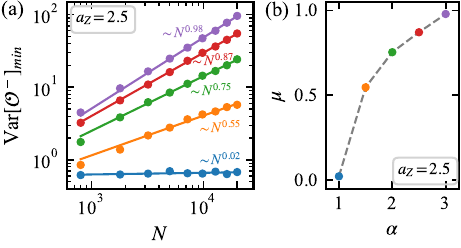}
\caption{Scaling of squeezing at fixed layer separation. (a) Scaling of minimal variance with system size for the $2d$ case at fixed layer separation $a_Z=2.5$ for different interaction exponents $\alpha$. Solid lines are fits to the function, $\text{Var}[\mathcal{O}^-]_{\rm min} \sim N^{\mu}$. (b) Change in scaling exponent $\mu$ for different $\alpha$, showing a fully collective regime ($\mu\approx 0$) for $\alpha = 1$ and non-collective regime ($\mu\approx 1$) for $\alpha = 3$, while a partially collective regime  ($0<\mu<1$) is seen for intermediate values of $\alpha$.
\label{fig:fig4}}
\end{figure}

So far we have considered the system size dependence of the minimal variance for the parameter $a_Z/L$ controlling the transition from fully collective to partially collective regimes. We now investigate the scaling of minimal variance with system size at fixed $a_Z$, which may be preferable in practice to retain fast time-scales (rather than the slow-down of squeezing times as $a_Z$ is scaled with $L$ weakening the interactions).  In Fig.~\ref{fig:fig4} (a) we plot the scaling of minimal variance with system size for different power-law exponents $\alpha$, represented by the color scheme of Fig.~\ref{fig:fig4} (b), for the $2d$ case. As expected, the dynamics remain in the fully collective regime for $\alpha\leq d/2$. For $\alpha\geq d/2$, with increasing $\alpha$ as the interactions get weaker, the minimal variance shows stronger system size dependence before ultimately becoming linear in the system size at $\alpha = 3$, entering the fully non-collective regime. For the intermediate power-law exponents, $1< \alpha < 3$, the sublinear scaling of minimal variance with system size denotes partially collective behavior. Thus the $\alpha=1$, and $\alpha =3$ cases represent the Heisenberg limited and standard quantum limited scaling of sensitivity, respectively, while for the intermediate $\alpha$ values there are still significant sensitivity gains beyond the standard quantum limit.

\section{Technical Details \& Extended Data}

\subsection{Identification of $(a_Z/L)_*$ from system size scaling exponent $p$}

\begin{figure}
\includegraphics[width=\columnwidth]{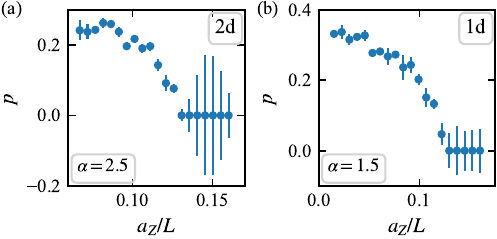}
\caption{Minimal variance system size scaling exponent $p$ versus $a_Z/L$ for (a) $2d$ and (b) $1d$ cases, where $\text{Var}[\mathcal{O}^-]_{\rm min} \sim N^{p}$. $p$ jumps from zero to a nonzero value at the critical $a_Z/L_*$.}
\label{fig:SI_MuVsaZL}
\end{figure}

In Fig 1(b) and (c) we identified the critical $(aZ /L)_{*}$ values marked by yellow markers via the scaling of the minimum variance with system size. We now show two examples of how the $(aZ /L)_{*}$ values are extracted from the scaling of minimal variance, $\text{Var}[\mathcal{O}^-]_{\rm min} \sim N^{p}$ in Fig.~\ref{fig:SI_MuVsaZL}. The advent of system size scaling at a critical $(a_Z/L)_*$ where $p$ jumps from zero to a nonzero value is clearly visible.

\subsection{Scaling of Variance Minima}
\begin{figure*}[hbt]
\includegraphics[width=\textwidth]{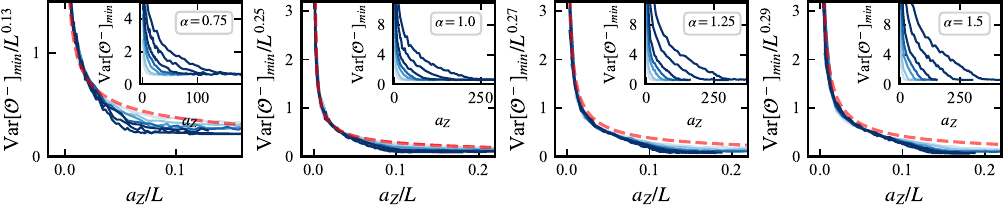}
\caption{Universal scaling collapse of minimal squeezing. Inset: Dependence of minimal variance $\text{Var}[\mathcal{O}^-]_{\rm min}$ on layer spacing $a_Z$, for different system sizes ($L = 100$ to 2500) for $\alpha =0.75$, 1, 1.25, and 1.5 in $1d$. Main panel: Rescaled minimal variance $\text{Var}[\mathcal{O}^-]_{\rm min}/L^{p \times d}$ versus aspect ratio $a_Z/L$ for the same system sizes, demonstrating a collapse onto a single curve. The red dashed lines $(a_Z/L)^{d_V}$ show fit to the scaling function with $d_V$ from Table.~\ref{tab:exponents}}
\label{fig:SI_VarVsaZL1d}
\end{figure*} 
\begin{figure*}[t]
\includegraphics[width=\textwidth]{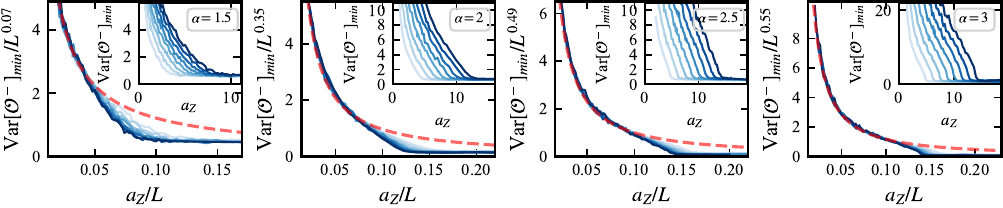}
\caption{Universal scaling collapse of minimal squeezing. Inset: Dependence of minimal variance $\text{Var}[\mathcal{O}^-]_{\rm min}$ on layer spacing $a_Z$, for different system sizes ($L = 30$ to 100) for $\alpha =1.5$, 2, and 2.5 in $2d$. Main panel: Rescaled minimal variance $\text{Var}[\mathcal{O}^-]_{\rm min}/L^{p \times d}$ versus aspect ratio $a_Z/L$ for the same system sizes, demonstrating a  collapse onto a single curve. The red dashed lines $(a_Z/L)^{d_V}$ show fit to the scaling function with $d_V$ from Table.~\ref{tab:exponents}}
\label{fig:SI_VarVsaZL2d}
\end{figure*} 
In the main text, Fig. 3 showed the scaling of variance minima for the representative case of $\alpha = 3$ in $2d$. Fig.~\ref{fig:SI_VarVsaZL1d} and Fig.~\ref{fig:SI_VarVsaZL2d} show the same for other power-law exponents $\alpha$ in $1d$  and $2d$ systems respectively.
In all cases, under the rescaling $\text{Var}[\mathcal{O}^-]_{\rm min}/L^{p \times d}$ the data collapses to a single curve, or scaling function, demonstrating that the aspect ratio $a_Z/L$ controls the transition, and that both distinct regimes show universal scaling of the minimal variance. Further, while finite size discrepancies near the critical value $(a_Z/L)_*$ are visible for smaller $L$, these are seen to vanish with increasing $L$. 
The red dashed line $(a_Z/L)^{d_V}$ is an excellent fit to the scaling function.


\subsection{Determining Critical Exponents}
To derive the exponents and associated uncertainties we follow the procedure proposed in Ref. \cite{Dresselhaus2022}, which defines a cost function to assess the best scaling collapse of the data. We first give a general description of the method, before describing its application to our work. We denote arbitrary `raw' data $\{y_{ij}, x_{ij} \}$ for the $j$'th point in the $i$'th data set (a given system size $N_i = L_i^d$). The data is ordered in increasing system size $\{N_1, N_2,...\}$. Consider the scaled data $\{\tilde{x}_{ij}, \tilde{y}_{ij} \} = \{ x_{ij} N_i^{d_{x}} , y_{ij} N_i^{d_{y}} \}$, for some exponents $d_{x}$ and $d_{y}$. Similarly, we define rescaled uncertainties $\{ \tilde{\delta}_{ij} \} = \{ \delta_{ij} N_i^{d_{y}} \}$. If the scaling collapse is successful, we expect $\tilde{y}_{ij}$ to be similar across different data sets for the same $\tilde{x}_{ij}$ values. In general however, such a comparison is difficult as different data sets will not have identical values of $\tilde{x}_{ij}$. We therefore use linear interpolation to generate curves $f_i(x)$ and $\delta_i(x)$ corresponding to $\tilde{y}_{ij}$ and $\tilde{\delta}_{ij}$, respectively, over a fine spatial grid $x$. To assess the success of the scaling collapse, we define the cost function \cite{Dresselhaus2022}
\begin{align}
   \lambda(d_x,d_y) = \frac{1}{\mathcal{N}} \sum_{i<k} \sum_x \frac{(f_i(x) - f_k(x))^2}{\delta_{i}(x)^2 + \delta_k(x)^2}, \label{eq:cost_function}
\end{align}
where the summation $x$ is over the entire (equally spaced) spatial grid for which data sets $i$ and $k$ both exist, and $\mathcal{N}$ is the total number of terms in the summations. The optimal value for the exponent is given by that which minimizes the cost function. While a good collapse is indicated by $\lambda_{\rm min} \leq 2 $, a poor collapse should have a much larger cost function \cite{Houdayer2004}. The exponent's uncertainty is then extracted from the range of values for $d_x$ and $d_y$ that give $\lambda(d_x,d_y) \leq  \lambda_{\rm min} +1 $ \cite{Houdayer2004}, where $\lambda_{\rm min}$ is the minimum of the cost function. This amounts to finding the range of exponents for which the cost function is within standard error of the minimum cost function.

\begin{table}[t]
(a) 
\begin{center}
\begin{tabular}{c|c|c|c}
  $\alpha$   & $d_{x}$ & $\Delta d_{x}$ & $\lambda_{\rm min}$  \\ \hline
  0.75   & -0.88   & 0.11 & 0.95   \\
    1  &   -0.9    & 0.08 & 1.83   \\
      1.25 & -0.92 & 0.05 & 0.95 \\
       1.5 & -0.93 & 0.04 & 0.37 \\
       2  & -0.97  & 0.04 & 1.78 
\end{tabular} \vspace{0.4cm}  
\end{center}
(b) 
\begin{center}
\begin{tabular}{c|c|c|c}
  $\alpha$   & $d_{x}$ & $\Delta d_{x}$ & $\lambda_{\rm min}$  \\ \hline
  1.5     & -0.8  & 0.26   & 0.92   \\
    2     &  -0.9 & 0.18   & 1.19  \\
      2.5 &  -0.87  & 0.11 & 0.78  \\
       3  &  -0.89  & 0.17 & 0.60 
\end{tabular}
\end{center}
\caption{Exponents $d_{x}$, uncertainties $\Delta d_{x}$, and minimum cost functions in (a) $1d$ and (b) $2d$.}
\label{tab:dx}
\end{table}

Throughout this analysis, we estimate the uncertainty $\delta_{ij}$ from the standard error of truncated Wigner samples. Intuitively, calculations within the collective manifold are expected to have less uncertainty.  In general, we find that the error scales proportional to ${\rm{Var}}[\mathcal{O}^-]_{\rm min} $ for a fixed system size. 
We estimate the numerical error for all $\alpha$ from the $\alpha = 3$ calculation, which yields $\delta \approx  0.04 {\rm{Var}}[\mathcal{O}^-]_{\rm min}$. This is the most non-collective system we consider in $2d$ and therefore represents a `worst-case' error. We have checked some specific  cases explicitly to confirm that this bounds the error for $\alpha < 3$, and that the error scales comparably in $1d$ . When evaluating Eq. \ref{eq:cost_function}, we typically ignore small system sizes, such as $L < 200 $ in $1d$ and $L < 30$ in $2d$, which are not expected to conform to the scaling hypothesis due to larger finite size effects. 

The results in Fig. 1 are used as a guide for selecting the appropriate range of $a_Z/L$, which implies a cutoff for $a_Z$ at a given $L$. We also need to avoid the range of $a_Z$ values between the collective and partially-collective phases. This is clearly visible in Fig. 2 of the main text where the data starts to deviated from the dotted (red) line. We therefore consider a range of $a_Z/L$ that is well inside the phase boundary indicated by Fig. 1.

\subsubsection{Scaling of critical $a_Z$ with $L$}
\begin{figure}[t]
\includegraphics[width=\columnwidth]{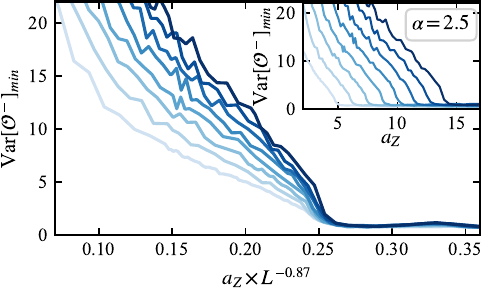}
\caption{Scaling collapse of the critical $a_Z$ value ($a_Z^*$) vs $L$. Inset: Dependence of minimal variance $\text{Var}[\mathcal{O}^-]_{\rm min}$ on layer spacing $a_Z$, for different system sizes ($L=$ 30 to 100) with $\alpha =2.5$ in $2d$. The opacity increases with system size. Main panel: $\text{Var}[\mathcal{O}^-]_{\rm min}$ vs $a_Z L^{-0.87}$, demonstrating a collapse of the critical values $a_Z^*L^{-0.87}$  to approximately a single point. 
}
\label{fig:xcollapse}
\end{figure}

We now apply this procedure to estimate the scaling of the critical $a_Z$ value, $a_Z^*$, with $L$.   We define the rescaled data $\{ \tilde{x}_{ij} , \tilde{y}_{ij} \} = \{ (a_Z)_j L_i^{d_{x}} , ({\rm{Var}}[\mathcal{O}^-]_{\rm min})_{ij} \} $ and only consider data with variance values in the range ${\rm{Var}}[\mathcal{O}^-]_{\rm min} \} \lessapprox \mathcal{O}(1)$ and $a_Z$ values in the vicinity of the transition.  
 Tables \ref{tab:dx} summarizes the results. Fig. \ref{fig:xcollapse} shows the rescaled data for the optimal $d_{x}$ value in the case of $\alpha = 2.5$ in $2d$.  In $1d$, where we have access to large system sizes, we find $d_x \in \{0.88,0.97\}$ in all cases, consistent with $d_x=1$ on the order of the uncertainty. The results are also broadly compatible with $a^*_z \sim L$ for the $2d$ case, although the uncertainty is larger. This justifies the use of $a_Z/L$ as the parameter controlling the phase and distance from the critical point in the main text. Larger deviations from $d_x = 1$ in $2d$ may be due to enhanced finite size effects. 

\subsubsection{Estimating exponent $p$ from variance minima.}
In this section we estimate the exponent $p$, where  $ \text{Var}[\mathcal{O}^-]  N^{-p}    =   y \big( a_{z} / L \big) $, from the variance minima data. In this instance the optimization is only carried out for a single exponent which scales the y-data. We apply the aforementioned optimization procedure with $\{ \tilde{x}_{ij} , \tilde{y}_{ij} \} = \{ (a_Z)_j/ L_i , ({\rm{Var}}[\mathcal{O}^-]_{\rm min})_{ij} N_i^{-p} \}.$  To restrict the data to the partially collective phase, we only consider data for which ${\rm{Var}}[\mathcal{O}^-]_{\rm min} $ is  $c$ times greater than  its value in the collective phase, i.e. ${\rm{Var}}[\mathcal{O}^-]_{\rm min} > c \lim_{a_Z \rightarrow \infty} {\rm{Var}}[\mathcal{O}^-]_{\rm min}$, where the limit is estimated by averaging over the variance in the collective phase. Typically we choose $c$ to be between 4 and 10, depending on $\alpha$. The results are listed in Table \ref{tab:pval}. Using these $p$ exponents, in Fig.~2 of main text, as well as in Figs.~\ref{fig:SI_VarVsaZL1d} and \ref{fig:SI_VarVsaZL2d}, we plot  $\text{Var}[\mathcal{O}^-]_{\rm min}L^{-p  d}$ versus aspect ratio $a_Z/L$ for different system size. The data in the partially collective phase collapses onto a single curve in all cases. Furthermore, the scaling function in this region is well described by $\text{Var}[\mathcal{O}^-]_{\rm min}L^{-p  d} = (a_Z/L)^{d_V}$, where $d_V$ is estimated in the subsequent section.

\begin{table}[t]
  (a) 
\begin{center}
  \begin{tabular}{c|c|c|c}
  $\alpha$   & $p$ & $\Delta p$ & $\lambda_{\rm min}^{(p)}$  \\ \hline
  0.75     & 0.13  & 0.06   & 0.89   \\
   1     &  0.25 &  0.06  & 1.30 \\
    1.25 &  0.27  & 0.06 & 0.96 \\
      1.5  & 0.29   & 0.06 & 1.49 \\
     2  &  0.35  & 0.05 & 0.57 \\   
\end{tabular} \vspace{0.4cm}  
\end{center}
 (b) 
\begin{center}
\begin{tabular}{c|c|c|c}
   $\alpha$   & $p$ & $\Delta p$ & $\lambda^{(p)}_{\rm min}$   \\ \hline
  1.5     &  0.035 & 0.055  & 0.50   \\
    2     &  0.175 & 0.055  &  0.63  \\
      2.5 &  0.245 & 0.065  & 0.48 \\
       3  &  0.275  & 0.065  &  0.61
\end{tabular}
\end{center}
\caption{Exponents $p$, uncertainties $\Delta p$, and minimum cost functions in (a) $1d$ and (b) $2d$, estimated from the minimum variance data (as in Fig. 2 of the main text).}
\label{tab:pval}
\end{table}

\subsubsection{Estimating exponents $d_V, d_{\tau},  \nu,\delta$.}
We now discuss the estimation of the four exponents $d_V, d_{\tau},  \nu,\delta$. Since $p$ has already been computed, we use $p = d_V(1- \delta)/d + \nu$ to constrain $\nu$ and reduce the number of unknowns to  three ($d_V$, $d_{\tau}$, $\delta$). 

\begin{figure*}[p]
\begin{center}
\subfloat[$\alpha = 0.75$, $N = 2500$]{
    \includegraphics[width = 7cm]{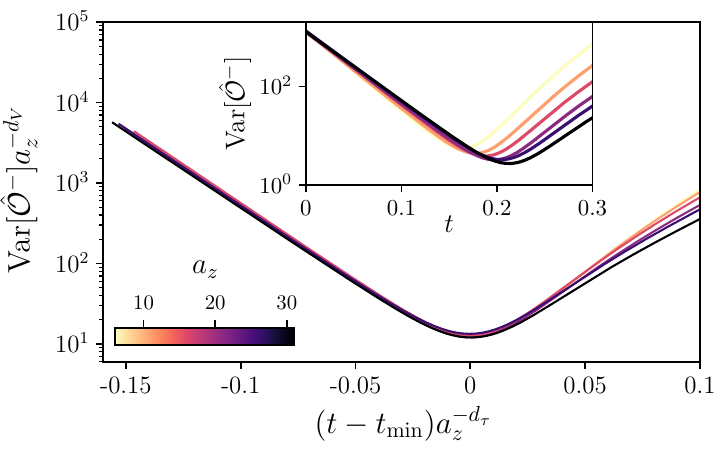} }
\subfloat[$\alpha = 0.75$]{
    \includegraphics[width = 7cm]{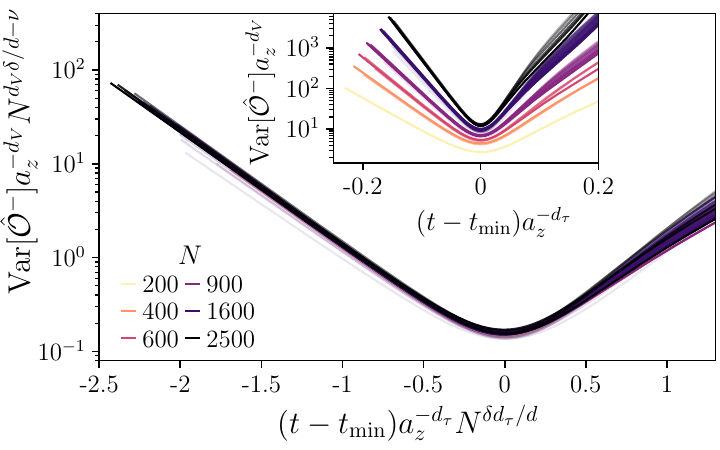} }

    \subfloat[$\alpha = 1$, $N = 2500$]{
    \includegraphics[width = 7cm]{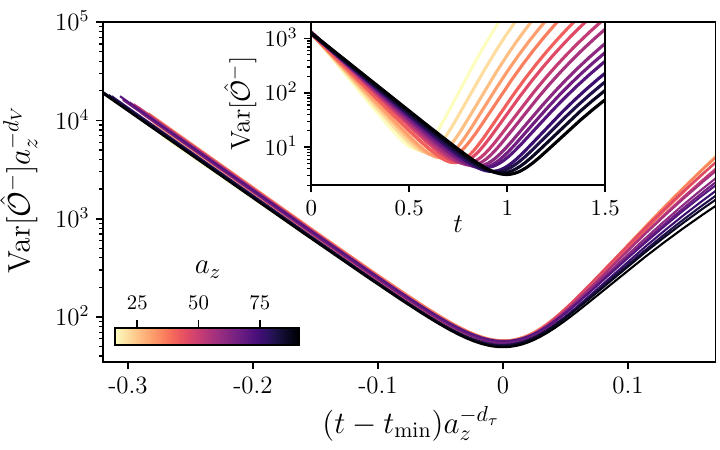} }
\subfloat[$\alpha = 1$]{
    \includegraphics[width = 7cm]{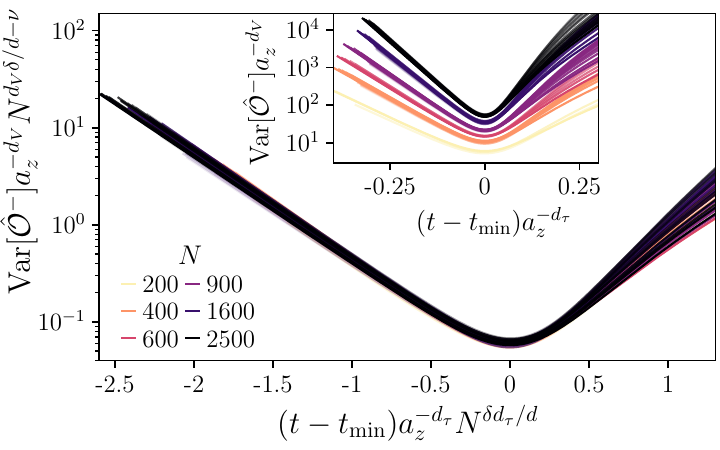} }

    \subfloat[$\alpha = 1.25$, $N = 2500$]{
    \includegraphics[width = 7cm]{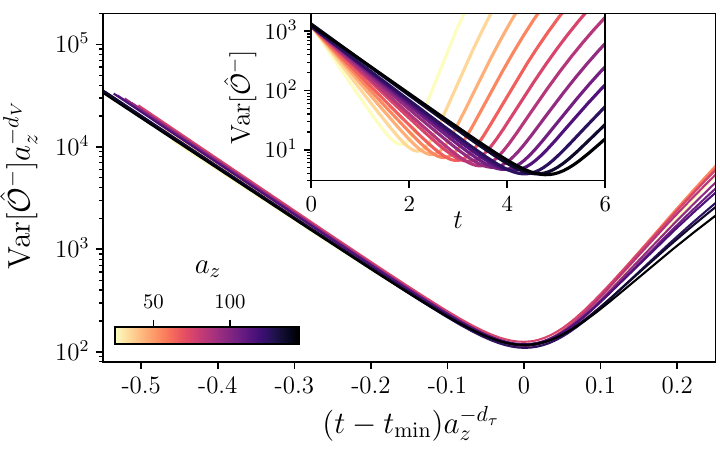} }
\subfloat[$\alpha = 1.25$]{
    \includegraphics[width = 7cm]{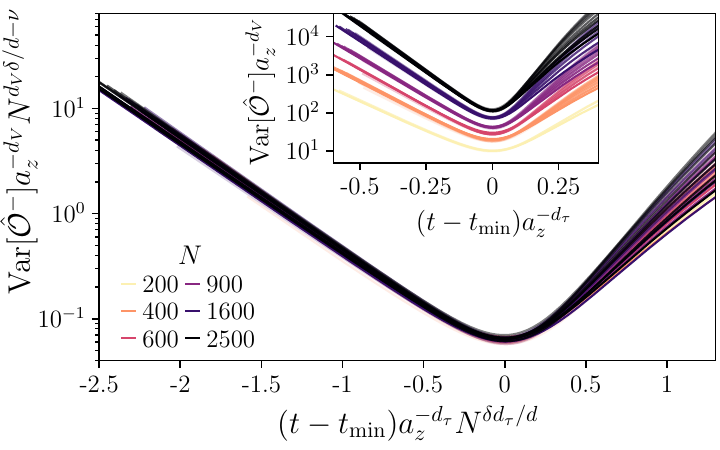} }

    \subfloat[$\alpha = 1.5$, $N = 2500$]{
    \includegraphics[width = 7cm]{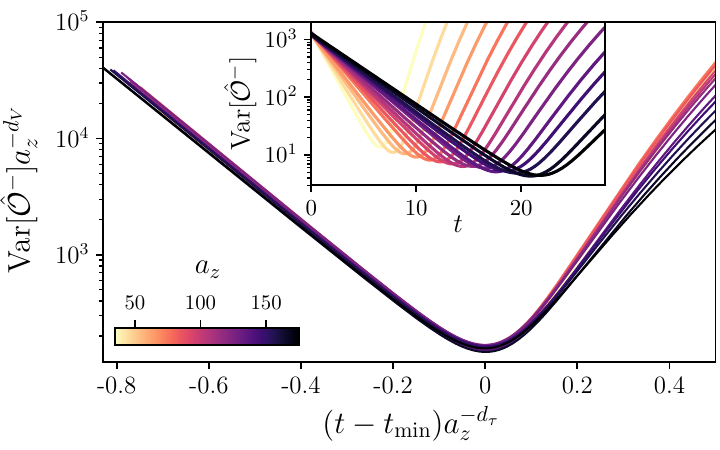} }
\subfloat[$\alpha = 1.5$]{
    \includegraphics[width = 7cm]{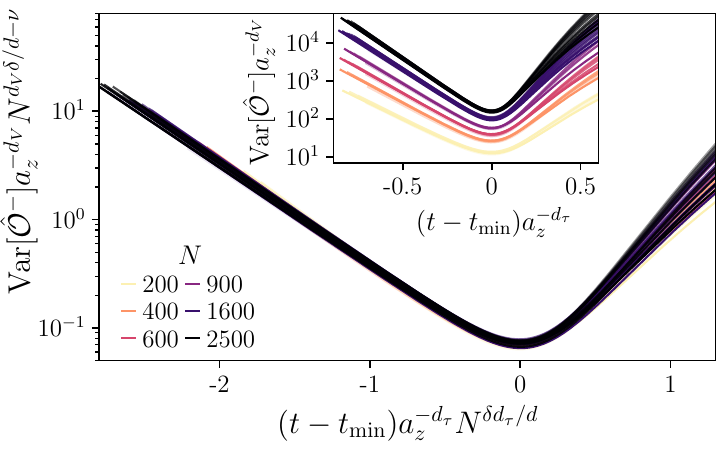} }
    \end{center}
    \caption{Raw data and scaling collapses for different $\alpha$ in $1d$, analogous to Fig.  4 of the main text albeit only plotting data that is used to extract the exponents.}
    \label{fig:SI_timecollapse_1d}
\end{figure*}

\begin{figure*}[p]
\begin{center}
\subfloat[$\alpha = 1.5$, $N = 10000$]{
    \includegraphics[width = 7cm]{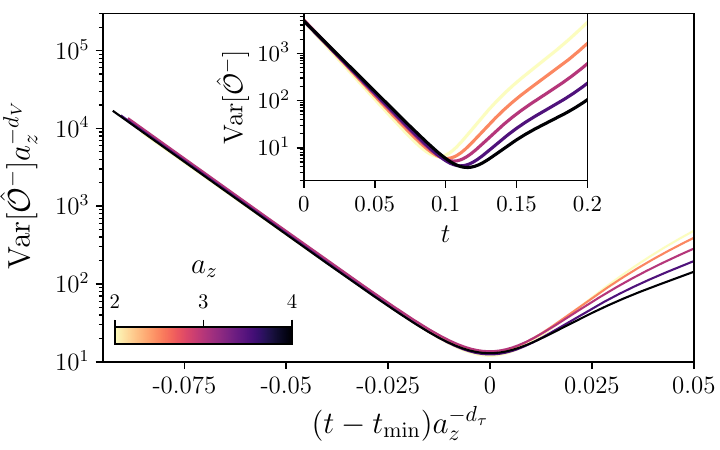} }
\subfloat[$\alpha = 1.5$]{
    \includegraphics[width = 7cm]{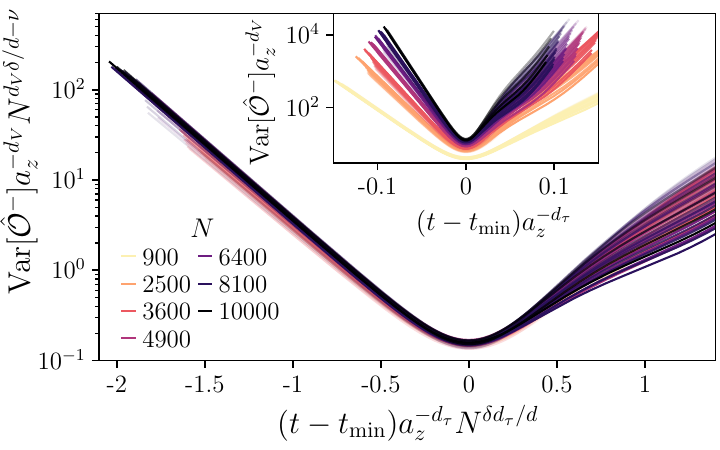} }

    \subfloat[$\alpha = 2$, $N = 10000$]{
    \includegraphics[width = 7cm]{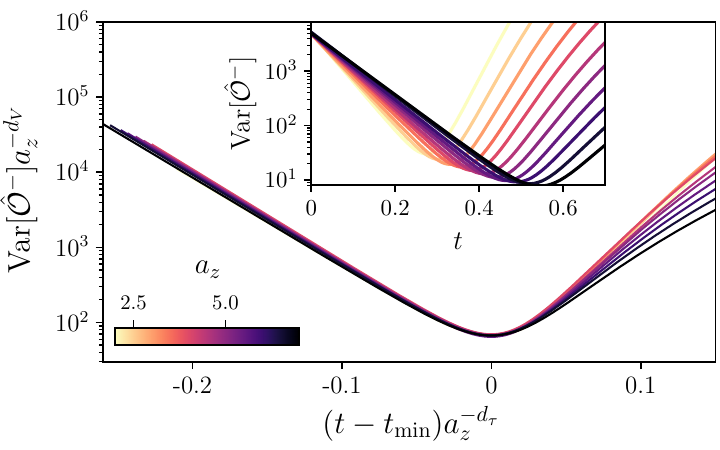} }
\subfloat[$\alpha = 2$]{
    \includegraphics[width = 7cm]{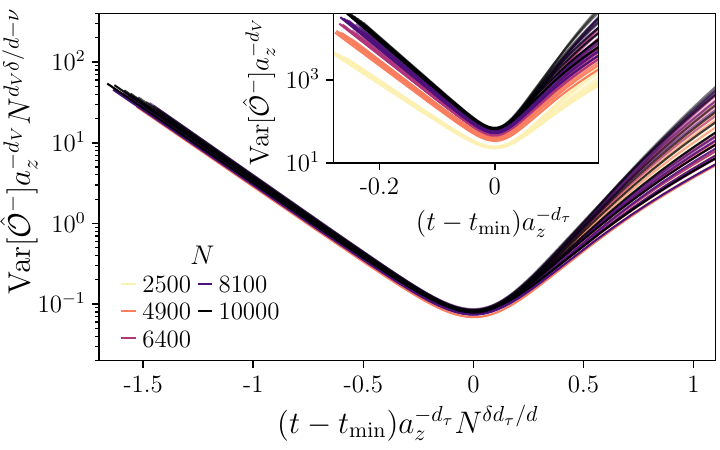} }

    \subfloat[$\alpha = 2.5$, $N = 8100$]{
    \includegraphics[width = 7cm]{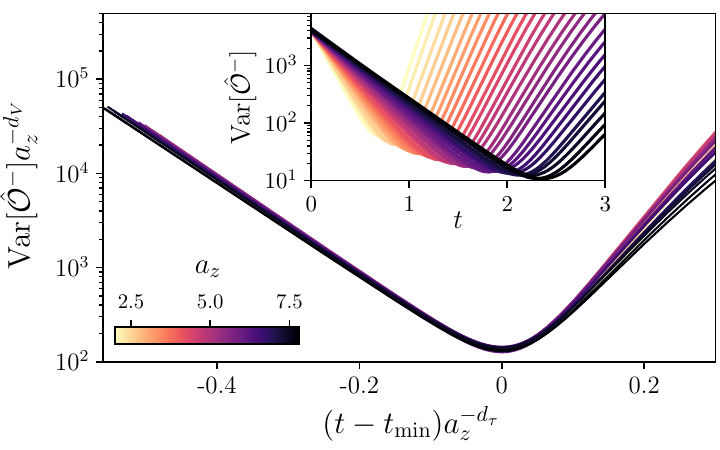} }
\subfloat[$\alpha = 2.5$]{
    \includegraphics[width = 7cm]{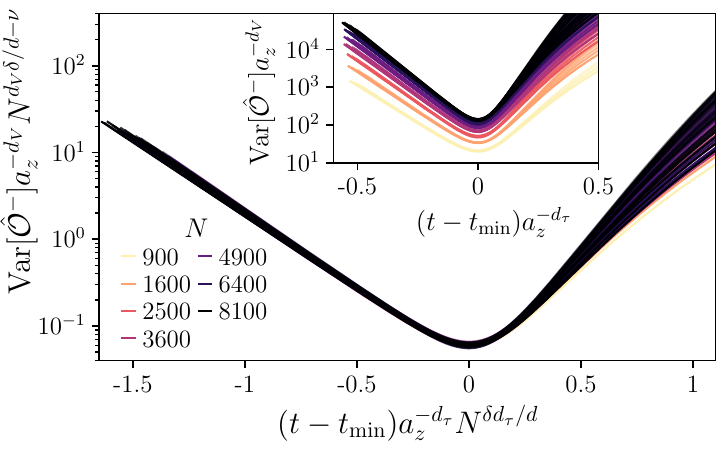} }

    \subfloat[$\alpha = 3$, $N = 10000$]{
    \includegraphics[width = 7cm]{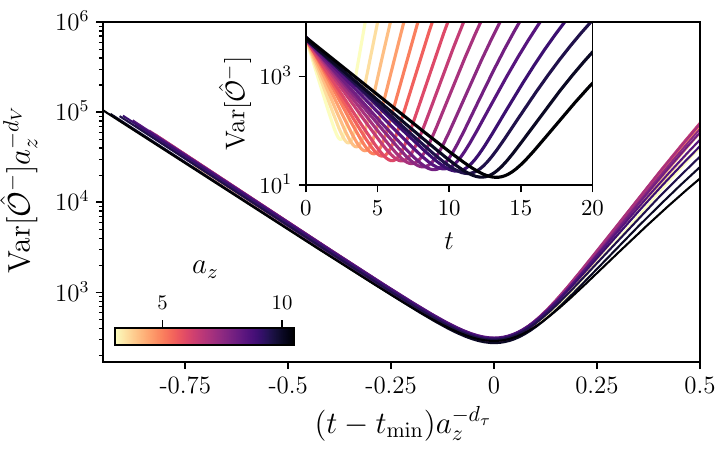} }
\subfloat[$\alpha = 3$]{
    \includegraphics[width = 7cm]{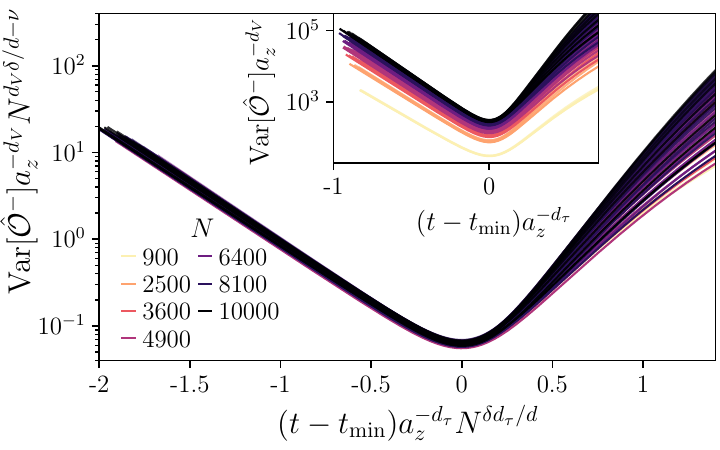} }
    \end{center}
    \caption{Raw data and scaling collapses for different $\alpha$ in $2d$, analogous to Fig.  4 of the main text.}
    \label{fig:SI_timecollapse_2d}
\end{figure*}

The exponents $d_V$ and $d_{\tau}$ are estimated simultaneously from the data for the largest available system size at a given $\alpha$: typically $N = 2500$ in $1d$ and $N = 10000$ ($L = 100$) in $2d$. As discussed, and shown in Fig. 4(a) of the main text,  this obeys the relation $
   \text{Var}[\mathcal{O}^-] a_Z^{-d_{V}}    = f_N [ (t - t_{\rm min})a_Z^{-d_{\tau}} ]. 
$ We therefore apply the optimization procedure over a two dimensional grid of $\{d_{V},d_{\tau}\}$ values using $\{ \tilde{x}_{ij} , \tilde{y}_{ij} \} = \{ (t - t_{\rm min})_j (a_Z)_i^{-d_{\tau}}, ({\rm{Var}}[\mathcal{O}^-]_{\rm min})_{ij} (a_Z)_i^{-d_{V}}  \} $. The results are shown in Table \ref{tab:exponents}. Fig. \ref{fig:SI_timecollapse_1d} ($1d$) and Fig. \ref{fig:SI_timecollapse_2d} ($2d$) include plots of $  \text{Var}[\mathcal{O}^-]a_Z^{-d_{V}} $ vs $(t - t_{\rm min})a_Z^{-d_{\tau}}$ for different $a_Z$ values,  as well as the corresponding raw data (inset). Data is only plotted over the range of $a_Z$ values that are used to extract the exponents.

Since the critical point occurs at finite $a_Z/L =\rm{constant}$, it is difficult to access a large range of $a_Z$ values in $2d.$ This issue is also exacerbated for lower $\alpha$ (even in $1d$) where this constant is reduced, as shown by the phase boundary in Fig. 1(a) of the main text. In these cases the range of $a_Z$ values in the partially collective phase that we can access becomes  small, and our results are less substantive. Nevertheless, we include these results for completeness. The reduced data range is reflected in larger relative uncertainties $\Delta d_V/d_V$ and $\Delta d_{\tau}/d_{\tau}$ for these cases. Note that we are still able to simulate a significant range of $N$ values in these cases, since $N = L^2$, which allows us to obtain reasonable estimates for the exponents $p$ and $\delta$.

To estimate $\delta$ we would like to  consider a similar procedure for the full scaling function in Eq. 2. To make the problem computationally tractable we assume the optimal values for $d_{\tau},$ $d_V$ and $p$. In this case the data is three dimensional and we use $t_j$ to index time, $(a_Z)_i$ for the layer spacing and $N_k$ for the system size. We therefore define
 $\{ \tilde{x}_{ijk} , \tilde{y}_{ijk} \} = \{ (t - t_{\rm min})_j (a_Z)_i^{-d_{\tau}}  N_k^{ \delta d_{\tau}/d} , ({\rm{Var}}[\mathcal{O}^-]_{\rm min})_{ij} (a_Z)_i^{-d_{V} } N^{d_V\delta/d-\nu} \}.$ The cost-function in Eq. \ref{eq:cost_function} is adapted to include a summation over the system size index, in addition to summing over rescaled time and all $(a_Z)_i$ curves. The cost-function is then evaluated for a range of $\delta$, with the results presented in Table \ref{tab:exponents}. Fig. \ref{fig:SI_timecollapse_1d} ($1d$) and Fig. \ref{fig:SI_timecollapse_2d} ($2d$) include plots of $\text{Var}[\mathcal{O}^-]  a_{z}^{-d_{V}}   N^{d_V\delta/d-\nu}$ vs $ (t - t_{\rm min}) a_{z}^{-d_{\tau}} N^{ \delta d_{\tau}/d}  $ for different $N$ and $a_Z$ values.  While the data typically does not conform to the criterion for a good fit $\lambda_{\rm min} \leq 2$, it is also generally not a poor fit, as would be indicated by a very large cost function $\lambda_{\rm min} \gg \mathcal{O} (1).$ 
In general, we find that the cost functions can be reduced by considering $a_Z/L$ values that are slightly further away from the transition. This lessens the impact of the cross-over region between the partially collective and collective regimes. However, this comes with the trade-off of reducing the range of data that can be considered, which tends to increase the uncertainty. Future works could improve on this by reaching larger system sizes, thereby expanding the range of accessible $a_Z$ values in the partially collective regime. 

 Since $d_V$, $d_{\tau}$, $\delta$ and $p$ are now known we can calculate $\nu = p - d_V(1- \delta)/d$, with its uncertainty following from error propagation. This is also presented in Table \ref{tab:exponents}. Since this calculation depends on the other exponents and their uncertainties, the uncertainty tends to be larger than in the aforementioned cases. 

\begin{table}[b]
(a)
\vspace{0.2cm}

\begin{tabular}{c|c|c|c|c|c|c|c|c|c|c}
   $\alpha$   & $d_V$ & $\Delta d_V$ & $d_{\tau}$ & $\Delta d_{\tau}$ & $\lambda_{\rm min}^{(d_V, d_{\rm \tau})}$ & $\delta$ & $\Delta \delta$ & $\lambda_{\rm min}^{(\delta)}$ & $\nu$ & $\Delta \nu$  \\ \hline
 0.75     &  -0.43       & 0.16 &  0.09 & 0.05  & 1.11 & 3.90 & 0.32 & 2.96 & -1.12 & 0.63 \\
   1     &  -0.61      &  0.16 & 0.25  & 0.05 & 3.31 & 1.07  & 0.12  & 3.04 & 0.21 & 0.08  \\
      1.25 &  -0.69      & 0.16 & 0.43  & 0.05 & 2.30 & 0.46  &  0.05 & 2.45 & 0.64  & 0.19  \\
       1.5 &  -0.69      & 0.2 & 0.63 & 0.07  & 1.72 &  0.25 & 0.04  & 2.11  &  0.81  &  0.29 
\end{tabular}

\vspace{0.4cm}
(b) 
\vspace{0.2cm}

\begin{tabular}{c|c|c|c|c|c|c|c|c|c|c}
   $\alpha$   & $d_V$ & $\Delta d_V$ & $d_{\tau}$ & $\Delta d_{\tau}$ & $\lambda_{\rm min}^{(d_V, d_{\rm \tau})}$ & $\delta$ & $\Delta \delta$ & $\lambda_{\rm min}^{(\delta)} $ &  $\nu$ & $\Delta \nu$    \\ \hline
  1.5     &   -0.88    &  0.34  & 0.16  &  0.08 & 1.14 &  4.2 & 0.3  & 6.00 & -1.37 &  0.70 \\
    2     &   -1.11    & 0.22  & 0.38  & 0.06   & 2.24 & 1.06  & 0.16  & 2.39 &  0.14 & 0.07 \\
      2.5 &   -1.22    & 0.22   & 0.7  &  0.08  &  1.27 & 0.34 & 0.08  & 1.19 & 0.65 & 0.23 \\
       3  &   -1.29     & 0.22  & 1.12 &  0.08 & 2.20 & 0.15 & 0.06 &  2.14 & 0.82 & 0.38
\end{tabular}
\caption{Exponents $d_V$, $d_{\tau},$ $\delta$, $\nu$ alongside their respective uncertainties, and the corresponding minimum cost functions $\lambda_{\rm min}^{(d_V, d_{\rm \tau})}$ and $\lambda_{\rm min}^{(\delta)}$ respectively. Data corresponds to (a) $1d$ and (b) $2d$.}
\label{tab:exponents}
\end{table}

\subsubsection{Different lattice structures}
Throughout this work we have primarily used $2d$ lattices with a square unit cell. Here, we explore the effect of different lattice structures on the exponents. Table \ref{tab:pval_geometry} and \ref{tab:exponents_geometry} show the critical exponents for $\alpha = 2$ in $2d$ with different lattice geometries: the square, triangular and hexagonal lattices. These are seen to be within error of each other, which is in keeping with the claim of universality since a change in the geometry of local interactions is a microscopic change.

\begin{table}[t]
\begin{center}
\begin{tabular}{c|c|c|c}
   {\rm Lattice}  & $p$ & $\Delta p$ & $\lambda^{(p)}_{\rm min}$   \\ \hline
   {\rm Square}       & 0.175 & 0.055  &  0.63  \\
   {\rm Triangular}     &  0.175  &  0.06  &  0.53  \\
    {\rm Hexagonal}       &  0.22 & 0.055  &  0.46
\end{tabular}
\end{center}
\caption{Exponents $p$, uncertainties $\Delta p$, and minimum cost functions in (a) $1d$ and (b) $2d$, estimated from the minimum variance data (as in Fig. 2 of the main text).}
\label{tab:pval_geometry}
\end{table}

\begin{table}[b]
\begin{tabular}{c|c|c|c|c|c|c|c|c|c|c}
   {\rm Lattice}  & $d_V$ & $\Delta d_V$ & $d_{\tau}$ & $\Delta d_{\tau}$ & $\lambda_{\rm min}^{(d_V, d_{\rm \tau})}$ & $\delta$ & $\Delta \delta$ & $\lambda_{\rm min}^{(\delta)} $ &  $\nu$ & $\Delta \nu$    \\ \hline
    Square     &   -1.11    & 0.22  & 0.38  & 0.06   & 2.24 & 1.06  & 0.16  & 2.39 &  0.14 & 0.07 
    \\ \hline
    Triangular     &   -1.11    & 0.24 & 0.38 & 0.07  & 2.22 & 1.04  & 0.16 & 3.28 & 0.15   & 0.07
    \\ \hline
    Hexagonal     &   -0.95    & 0.18  & 0.35 &  0.06  &  1.48  & 1.06 & 0.16 & 1.38  & 0.19 &  0.07
\end{tabular}
\caption{Exponents $d_V$, $d_{\tau},$ $\delta$, $\nu$ alongside their respective uncertainties, and the corresponding minimum cost functions $\lambda_{\rm min}^{(d_V, d_{\rm \tau})}$ and $\lambda_{\rm min}^{(\delta)}$ respectively. Data corresponds to $2d$ with $\alpha = 2$, comparing different lattice structures.}
\label{tab:exponents_geometry}
\end{table}


\bibliographystyle{apsrev4-2}
\bibliography{TMS_new}